 \documentclass[article, 11pt, onecolumn]{IEEEtran}

\usepackage{cite}
\usepackage{array}
\usepackage{array}
\usepackage{glo}
\usepackage{epsfig}
\usepackage{rotating}
\usepackage{latexsym}
\usepackage{color}
\usepackage{pstricks}
\usepackage{times}
\usepackage{changebar}
\usepackage{pifont}
\usepackage{amssymb}
\usepackage{amsmath}
\usepackage{bm}
\usepackage[english]{babel}
\usepackage[latin1]{inputenc}
\usepackage{ifthen}
\usepackage{fixltx2e}

\usepackage{hyperref}

\input epsf

\newcommand\comments[1]{}

\newtheorem{theorem}{Theorem}
\newtheorem{lemma}{Lemma}

\newtheorem{proposition}{Proposition}

\newtheorem{remark}{Remark}

\begin{document}

\setlength{\changebarsep}{ 0mm}
\setlength{\changebarwidth}{3mm}
\setcounter{changebargrey}{60}

\bibliographystyle{ieeetr}
\setlength{\evensidemargin}{-0.309in}
\setlength{\oddsidemargin}{-0.309in}
\setlength{\voffset}{-0.55in}
\setlength{\textheight}{9.68in}
\setlength{\topmargin}{0.01in}
\setlength{\headheight}{0in}
\setlength{\headsep}{0in}
\setcounter{page}{1}
\date{}

\author{\large Wei Liu,   Ruoyu Sun, Zhi-Quan Luo and  Jiandong Li\\
\thanks{*This work was performed while the first author was a visiting scholar at University of Minnesota.  }%
\thanks{W. Liu and J. Li are with  State Key Labs of ISN, Xidian University, 710071,
Xian, Shaanxi, P.R. China. Email:\{liuweixd, jdli\}@mail.xidian.edu.cn.}%
\thanks{R. Sun (corresponding author) is with the Department of Management Science and Engineering, Stanford University, USA.
Email: ruoyus@stanford.edu. This work was done while this author was a PhD student in the department of ECE, University of Minnesota.}%
\thanks{Z. Q. Luo is with the Department of Electrical and Computer Engineering, University of Minnesota, 200 Union Street
SE, Minneapolis, MN 55455, USA. Email:luozq@ece.umn.edu.}%
\thanks{
 Part of this paper is published in ICASSP 2014 \cite{Ruoyu2014icassp}.  }
 }

 \hbadness =
10000

\title{Globally Optimal Joint Uplink Base Station Association and Beamforming*}

\maketitle

\thispagestyle{empty}

\pagestyle{plain}

\begin{abstract}
The joint base station (BS) association and beamforming problem has been studied  extensively in recent years,
yet the computational complexity for even the simplest SISO case has not been fully characterized.
 In this paper, we consider the problems for
 an uplink SISO/SIMO cellular network under the max-min fairness criterion.
We first prove that the problems for both the SISO and SIMO scenarios are polynomial time solvable.
Secondly, we present a fixed point based binary search (BS-FP) algorithm for both SISO and SIMO scenarios whereby a QoS (Quality of Service) constrained subproblem is solved at each step by a fixed point method.
Thirdly, we propose a normalized fixed point (NFP) iterative algorithm to directly solve the original problem and prove its geometric convergence to  global optima. Although it is not known whether the NFP algorithm is a polynomial time algorithm,
empirically it converges to the global optima orders of magnitude faster than the polynomial time algorithms,
making it suitable for applications in huge-scale networks.
\end{abstract}



\section{Introduction}
To meet the surging mobile traffic demand, wireless cellular networks have increasingly relied
on low power transmit nodes such as pico base stations (BS) to work in concert with the existing macro BSs.
Such a heterogeneous network (HetNet) architecture can provide substantially improved data service to cell edge users.

One crucial problem in the system design of future networks is how to associate mobile users with serving BSs.
The conventional greedy scheme that associates receivers with the transmitter providing the strongest signal and its modern variant Range Extension \cite{LTEHetNet} may be suboptimal during periods of congestion.
 A more systematic approach is to jointly design BS association and other system parameters so as to maximize a network-wide utility.
The early work in this direction \cite{Yates95} and \cite{Hanly1995} proposed a fixed point iteration to jointly adjust BS association and power allocation in the uplink (UL), while subject to QoS (Quality of Service) constraints, with the goal to minimize total transmit power.
 The global convergence of this algorithm has been established when the problem is feasible.
 This algorithm has been extended to the single input single output (SISO)  cellular network with power budget constraints for both the UL
 and the downlink (DL) in \cite{Yates1997, RFLiuDL}, respectively.    The fixed point algorithm in \cite{Yates95, Hanly1995} can also be interpreted as an
  alternating optimization approach: fix the BS association, each user updates it power to satisfy the QoS Constraint; fix the power,
  each user updates its BS association to maximize its SINR (Signal to Interference plus Noise Ratio).  This alternating optimization approach
   was extended to joint BS association, power allocation and beamforming for UL SIMO cellular network in \cite{Farrokh-Rashid-Farrokhi}.

Recently, various approaches have been applied to tackle the BS association problem \cite{Madan2010,Sparse2013, Sun2013,Ye2012,YuWei2013}.
The work in \cite{Madan2010} proposed to solve a utility maximization problem by alternately optimizing over BS association and other system parameters
for SISO DL cellular network.
References \cite{Sparse2013, Sun2013} considered a partial CoMP (Coordinated Multiple Point) transmission strategy, i.e., allowing one user to be served by
multiple BSs for MIMO DL cellular network. 
They proposed sparse optimization techniques to compute a desirable BS association that incurs low overhead.
References \cite{Ye2012,YuWei2013} studied the joint design of BS association and frequency resource allocation for a fixed transmission power for DL SISO Cellular Network .






The computational complexity of maximizing a certain utility function by joint BS association and power allocation has been studied in different scenarios \cite{Hong2012,Maz2012,Sun2012}.
For the sum rate utility function, the NP hardness of the joint design problem has been established for both the UL MIMO cellular network \cite{Hong2012}
and the DL MIMO cellular network \cite{Maz2012}.
As a counterpart, for the max-min fairness utility, the joint design problem in a DL SISO network is shown to be NP-hard in general,
while for some special cases (with an equal number of users and BSs and under additional QoS constraints) this problem is shown to be polynomial time solvable \cite{Ruoyu2014jsac}.

Despite the extensive research, a fundamental theoretical question remains open: is the joint design problem under max-min fairness criterion in an UL cellular network polynomial time solvable? In addition, to handle large scale networks, fast algorithms with performance guarantee (not only theoretically polynomial time solvable) are much needed.


In this paper, we aim to resolve these issues. In particular, we consider the joint BS association and beamforming  problem  under the
the max-min fairness criterion for both  UL SISO and SIMO cellular networks.  Our main contributions are as follows:
\begin{enumerate}
\item We prove that the problems for both SISO and SIMO scenarios are polynomial time solvable.
To be more specific, we show that the problem for SIMO (resp. SISO) networks can be solved by a binary search method whereby each subproblem can be solved by SDP (resp. LP), and we refer to this algorithm as BS-SDP (resp. BS-LP).  This is rather surprising since the considered optimization problem involves discrete variables and
falls into the class of mixed integer nonlinear programming.

\item We present a globally optimal fixed point based binary search (BS-FP) algorithm  in which the QoS constrained subproblems are solved by fixed point  algorithms, which can avoid invoking  a LP solver for SISO scenarios or a SDP solver for SIMO scenarios.

\item We propose a  normalized fixed point (NFP) algorithm to directly solve the joint BS association and beamforming max-min fairness problem, which can avoid binary search.
Theoretically, using results from the concave Perron-Frobenius theory \cite{krause2001concave,krause1994relative}, we prove the geometric convergence of the proposed algorithm to global optima. At this point, we are only able to show the pseudo-polynomial time, not polynomial time complexity, of the NFP algorithm, although empirically it is much faster than the polynomial time algorithms BS-LP and BS-SDP.  In fact, the NFP algorithm converges in less than 20 iterations for networks with hundreds of BSs and users, as shown in the numerical experiments.
 \end{enumerate}

We summarize the computational complexity results for the joint BS association and beamforming problem under the max-min fairness criterion for both  UL and DL cellular networks in
Table \ref{table of comeplexity of UL_and_DL_max_min} below.
\begin{table}[htbp]\label{table of comeplexity of UL_and_DL_max_min}
\caption{The Complexity Status of the Joint BS Association and Beamforming Problem  (max-min fairness)}
\centering
\begin{tabular}{|c|c|c|c|}
\hline
 &\multicolumn{1}{c|}{Fixed BS association}&\multicolumn{2}{c|}{Joint}\\
\hline
& &{UL}& {DL}\\
\hline
SISO &Polynomial \cite{Luo2008}&Polynomial (Theorem 1)&NP-Hard \cite{Ruoyu2014jsac}\\
\hline
SIMO&Polynomial  \cite{Ya-Feng-Liu-1}&Polynomial (Theorem 3)&NP-Hard \cite{Ruoyu2014jsac}\\
\hline
 MISO&Polynomial  \cite{bengtsson2001optimal,wiesel2006linear}&Unknown\footnotemark&NP-Hard \cite{Ruoyu2014jsac}\\
\hline
MIMO&NP-Hard \cite{Ya-Feng-Liu-2}\footnotemark& Unknown& NP-Hard \cite{Ruoyu2014jsac}\\
\hline
\end{tabular}
\end{table}
\footnotetext[1]{The SIMO UL network (multiple antennas at each user while single antenna at each BS) does not seem to be a very interesting scenario.}

\footnotetext[2]{When the number of antennas at each transmitter (resp. receiver) is at least three and the number of antennas at each receiver (resp. transmitter) is at least two.}

This paper is structured as follows.   In Section \ref{sec:System Overview_miso_max_min}, the max-min fairness problem by joint BS association, power control and beamforming problem for an UL SIMO cellular network is introduced.   In Section
\ref{sec:Joint_BS_Association_and_Power_Control_for_SISO_System}, we investigate the SISO scenario. We first prove the polynomial time solvability for  the SISO scenario, then present  a BS-FP algorithm and finally propose the NFP algorithm.  In Section \ref{sec:Joint_BS_Association_and_Beamforming_for_SIMO_System}  we investigate the SIMO scenario. We first prove the polynomial time solvability for  the SIMO scenario, then present  a BS-FP algorithm and finally propose the NFP algorithm.   Simulation results are provided in Section \ref{sec:Simulation_Results_miso_max_min} to compare the efficiency and the effectiveness of the algorithms.  Finally, some concluding remarks are offered
in Section \ref{sec:Conclusions_miso_max_min}.

\section{System Overview}\label{sec:System Overview_miso_max_min}
Consider an uplink cellular network where $K$ mobile users transmit to $N$ BSs. Each user is  equipped with a single antenna and the $n$th BS is
 equipped with $M_{n}\geq 1$ antennas, $n=1,\cdots,N$. They share the same time/frequency resource for transmission. Each user is to be associated with exactly one BS, but one BS can serve multiple users.
 Assuming  that the transmitted signal from the $k$th mobile user is $s_{k}$, the received signal $\pmb{y}_{n}$ at the $n$th BS may be expressed as
 \begin{align}\label{eq:uplink_assocation_miso_sysyem_model_1}
\pmb{y}_{n}=\sum_{k=1}^{K}\pmb{h}_{nk}s_{k}+\pmb{n}_{n},
\end{align}
where the $M_{n}$-dimensional vector $\pmb{h}_{nk}$ denotes the flat fading channel coefficient between the $k$th mobile user and the $n$th BS, while the $M_{n}$-dimensional vector $\pmb{n}_{n}$  denotes the AWGN with zero mean and a covariance matrix of $\sigma_{n}^{2}\pmb{I}$.
Let $\bm a= (a_1,a_2,\dots, a_K)$ denote the association profile, i.e.,
$a_k = i$ if user $k$ is associated with BS $i$.
For the $k$th user, the BS $a_{k}$ invokes a $M$-dimensional unit-norm linear receiver $\pmb{u}_{a_{k},k}$ to generate the decision signal $\tilde{s}_{k}$ for the
 $k$th user as
 \begin{align}\label{eq:uplink_assocation_miso_sysyem_model_2}
\tilde{s}_{k}&=\pmb{u}_{a_{k},k}^{H}\pmb{y}_{a_{k}}=\pmb{u}_{a_{k},k}^{H}\sum_{j=1}^{K}\pmb{h}_{a_{k},j}s_{j}+\pmb{u}_{i,k}^{H}\pmb{n}_{k}\nonumber\\
&=\pmb{u}_{a_{k},k}^{H}\pmb{h}_{a_{k},k}s_{k}+\pmb{u}_{a_{k},k}^{H}\sum_{j=1,j\not=k}^{K}\pmb{h}_{a_{k},j}s_{j}+\pmb{u}_{a_{k},k}^{H}\pmb{n}_{a_{k}}.
\end{align}

The SINR for the $k$th user is given by
  \begin{align}\label{eq:uplink_assocation_miso_sysyem_model_3}
\textrm{SINR}_{k}=\frac{p_{k}\pmb{u}_{a_{k},k}^{H}\pmb{h}_{a_{k},k}\pmb{h}_{a_{k},k}^{H}\pmb{u}_{a_{k},k}}{\sigma_{a_{k}}^{2}+\pmb{u}_{a_{k},k}^{H}\sum_{j=1,j\not=k}^{K}p_{j}\pmb{h}_{a_{k},j}\pmb{h}_{a_{k},j}^{H}\pmb{u}_{a_{k},k}},
\end{align}
where $p_{k}=E[s_{k}s_{k}^{*}]$ denotes the transmission power of the $k$th user.

 \begin{equation}\label{max min}
\begin{split}
\textrm{(P)}: \max_{\substack{\pmb{a},\pmb{p},\\ \{\pmb{u}_{n,k}\}_{\substack{ k=1,\dots,K,\\ n=1,\dots,N}}}}  &  \min_{k=1,\dots, K} \textrm{SINR}_{k}\\
\textrm{s.t.} \quad  & 0 \leq p_k \leq \bar{p}_k, \;  k=1,\dots,K, \\
            &  a_k \in \{1,2,\dots, N \}, \;  k=1,\dots,K, \\
            &  \|\pmb{u}_{n,k}\|=1, \; k=1,\dots,K,n=1,\dots,N,
\end{split}
\end{equation}
where $\bar{p}_{k}$ is the power budget of the $k$th user.

In the following, we will investigate the SISO and SIMO scenarios, respectively.   Note that the results and algorithms for SIMO are more general than
those for SISO; but to make the ideas easier to understand, we will present those for SISO separately.

\section{Joint BS Association and Power Control for UL SISO Celluar Networks}\label{sec:Joint_BS_Association_and_Power_Control_for_SISO_System}
In this section, we consider the SISO system where $M_{n}=1,\forall \ n$.  In this case,  each beamforming vector ${\pmb{u}_{n,k}}$ reduces to a scaler $u_{n,k}$ and the optimal $u_{n,k}$  is given by $\hat{u}_{n,k}=h_{n,k}^{\dagger}/\|h_{n,k}\|$, where the superscript $\dagger$ denotes the complex conjugation.  Substituting $\hat{u}_{n,k}$ into problem  (P) yileds
\begin{equation}\label{max min}
\begin{split}
\textrm{(P\textsubscript{SISO})}: \max_{\bm p,\bm a  }  & \min_{k=1,\dots,K} {\rm SINR}_k = \frac{  g_{a_k k} p_{k} }{\sigma_{a_k}^2 + \sum_{j \neq k} g_{a_k j}p_j  }, \\
\textrm{s.t.} \quad  & 0 \leq p_k \leq \bar{p}_k, \;  k=1,\dots,K, \\
            &  a_k \in \{1,2,\dots, N \},  \; k=1,\dots,K,
\end{split}
\end{equation}
where $g_{ik}=\|h_{ik}\|^{2}$ is the channel gain between user $k$ and BS $i$.

Optimizing $\bm p$ and $\bm a$ separately is easy. Specifically, given a fixed BS association $\bm a$, the formulation (\ref{max min}) is a max-min fairness power control problem for an I-MAC (interfering multiple-access channel).
It can be solved in polynomial time using a binary search strategy whereby a QoS constrained subproblem is solved by LP (Linear Programming) at each step \cite{Luo2008}. Moreover, notice that  the interference for user $k$ is $\sum_{j \neq k} g_{a_k j}p_j$, which only depends on $a_k$, and does not depend on $a_j, \forall j\neq k$.  Thus, given a power vector $\bm p$, the optimal association of each user $k$ does not depend on the choices of other users and can be easily computed:
 \begin{equation}\label{optimal a, formula}
 a_k = \arg \max_{ n\in \{1,\dots, N \} } \left\{ \frac{  g_{n k} p_{k} }{\sigma_{n}^2 + \sum_{j \neq k} g_{n j}p_j  } \right\}.
 \end{equation}
In case of multiple $n$'s that achieve the maximum in (\ref{optimal a, formula}),  we just use  $\arg \max \{\dots\}$ to represent any element achieving the maximum.
However, it is not straightforward to jointly optimize the continuous variable $\bm p$ and the discrete variable $\bm a$, and this is the focus of our work.

\subsection{Polynomial Time Solvability}\label{sec:System Overview_MIMO_OSTBC_Optimal_Relay}
In this section, we will prove that the problem  (P\textsubscript{SISO}) is polynomial time solvable.

\begin{theorem}\label{thm_polynomial_time_solvable}
The  problem  \textrm{(P\textsubscript{SISO})},  i.e.,  maximizing the minimum  SINR by joint BS association and power control for an uplink SISO cellular network, is polynomial time solvable.
\end{theorem}

\begin{IEEEproof}[Proof of Theorem \ref{thm_polynomial_time_solvable}]
The max-min fairness problem is closely related to the QoS constrained problem, i.e., minimize the total transmission power subject to the QoS constraints.
The QoS constrained joint BS association and power allocation problem is given as follows:
\begin{subequations}
\begin{align}
\textrm{(P\textsubscript{SISO-QoS})}:\min_{\pmb{p},\pmb{a} }  & \sum_{k=1}^K p_k,\nonumber \\
\textrm{s.t.} \quad  & 0 \leq p_k \leq \bar{p}_k, \; k=1,\dots,K, \label{min power_SISO_1} \\
            &   a_k \in \{1,2,\dots, N \}, \; k=1,\dots,K,\label{min power_SISO_2} \\
            & {\rm SINR}_k = \frac{g_{a_k k} p_{k} }{\sigma_{a_k}^2 + \sum_{j\neq k}g_{a_k j}p_j  } \geq \gamma, \; \label{min power_SISO_3} \\
            & k=1,\dots, K,\nonumber
\end{align}
\end{subequations}
where $\gamma$ is the required SINR value.  Thus, problem \textrm{(P\textsubscript{SISO})}  can be solved by a sequence of subproblems of the form \textrm{(P\textsubscript{SISO-QoS})} and a binary search on $\gamma$.   In the following, we will show that the QoS constrained subproblem \textrm{(P\textsubscript{SISO-QoS})} can be transformed to a linear programming (LP).

Since the optimal BS association $a_{k}$ is given by  (\ref{optimal a, formula}), we have
\begin{align}
& \exists \ \pmb{a}~~ \textrm{satisfying}~~ \textrm{(\ref{min power_SISO_2}) and (\ref{min power_SISO_3})} \nonumber\\
 &\Longleftrightarrow
\max_{n \in \{1,\dots, N\}} \frac{g_{nk}p_{k}}{\sigma_{n}^{2}+\sum_{j\not=k} g_{nj}p_{j}}\geq \gamma,\  k=1,\dots,K\\
&\Longleftrightarrow
p_{k} \geq \min_{n \in \{1,\dots, N\}}\gamma  (\pmb{g}_{k}^{n}\pmb{p}+\tilde{\sigma}_{nk}^{2}),\  k=1,\dots,K,
\end{align}
where $\pmb{g}_{k}^{n}=[g_{n1}/g_{nk},\cdots,g_{n(k-1)}/g_{nk},0,g_{n(k+1)}/g_{nk},\cdots,$ $g_{nK}/g_{nk}]$ and $\tilde{\sigma}_{nk}^{2}=\sigma_{n}^{2}/g_{nk}$.
Consequently,  the problem (P\textsubscript{SISO-QoS}) is equivalent to the following problem:
 \begin{equation}\label{min power_SISO_4}
\begin{split}
\textrm{(P\textsubscript{SISO-QoS-1})}:\min_{\pmb{p} }  & \sum_{k=1}^K p_k, \\
\textrm{s.t.} \quad  & 0 \leq p_k \leq \bar{p}_k, \; k=1,\dots,K, \\
            &  p_{k} \geq \min_{n}\gamma  (\pmb{g}_{k}^{n}\pmb{p}+\tilde{\sigma}_{n}^{2}), \; k=1,\dots,K.
\end{split}
\end{equation}

According to [4, Lemma 4(2)], the equation
\begin{equation}\label{min power_SISO_6}
p_{k} =\min_{n}\gamma  (\pmb{g}_{k}^{n}\pmb{p}+\tilde{\sigma}_{n}^{2}).
\end{equation}
  has a unique fixed point, denoted as  $ \hat{\bm p} $.

Let us consider another QoS constrained problem (P\textsubscript{SISO-QoS-2}) below:
 \begin{equation}\label{min power_SISO_5}
\begin{split}
\textrm{(P\textsubscript{SISO-QoS-2})}:\max_{\pmb{p} }  & \sum_{k=1}^K p_k, \\
\textrm{s.t.}    \quad  & 0 \leq p_k \leq \bar{p}_k, \ k=1,\dots, K;  \\ 
            &  p_{k} \leq \min_{n}\gamma  (\pmb{g}_{k}^{n}\pmb{p}+\tilde{\sigma}_{n}^{2}), \; k=1,\dots,K.
\end{split}
\end{equation}
This problem is always feasible since $(0,0,\dots, 0)$ is one feasible solution; in addition, the objective value is upper bounded by $\sum_k \bar{p}_k$,
thus the optimal solution to \textrm{(P\textsubscript{SISO-QoS-2})} always exists.

The following result shows that   \textrm{(P\textsubscript{SISO-QoS-1})} and \textrm{(P\textsubscript{SISO-QoS-2})} are ``equivalent''.
\begin{lemma}\label{lemma of equivalence of SISO1 and SISO2}
The two problems \textrm{(P\textsubscript{SISO-QoS-1})} and \textrm{(P\textsubscript{SISO-QoS-2})} are equivalent in the following sense:
if \textrm{(P\textsubscript{SISO-QoS-1})} is infeasible, then for any optimal solution to \textrm{(P\textsubscript{SISO-QoS-2})},
denoted as $\tilde{\bm p}$, there exists some $k$ such that $ \tilde{p}_k < \min_n \gamma (\bm g_k^n \tilde{\bm p} + \tilde{\sigma}^2_n ) $;
if \textrm{(P\textsubscript{SISO-QoS-1})} is feasible, then $\hat{\bm p} $  is the unique optimal solution to both \textrm{(P\textsubscript{SISO-QoS-1})}
and  \textrm{(P\textsubscript{SISO-QoS-2})}, where $ \hat{\bm p} $ is the unique fixed point of \eqref{min power_SISO_6}.
\end{lemma}

\emph{Proof}: If \textrm{(P\textsubscript{SISO-QoS-1})} is infeasible, suppose $\tilde{\bm p}$ is one optimal solution to \textrm{(P\textsubscript{SISO-QoS-2})}, then  $ \tilde{p}_k \leq \min_n \gamma (\bm g_k^n \tilde{\bm p} + \tilde{\sigma}^2_n ), \forall k $.
 We must have $ \tilde{p}_k < \min_n \gamma (\bm g_k^n \tilde{\bm p} + \tilde{\sigma}^2_n ) $
for some $k$; otherwise $ \tilde{p}_k = \min_n \gamma (\bm g_k^n \tilde{\bm p} + \tilde{\sigma}^2_n ), \forall k $, implying that
$\tilde{\bm p}$ is a feasible solution to \textrm{(P\textsubscript{SISO-QoS-1})}, a contradiction.

If \textrm{(P\textsubscript{SISO-QoS-1})} is feasible, we claim that its optimal solution $\bm p^*$ must satisfy \eqref{min power_SISO_6}.
In fact, if one $p_{k}^* >\min_{n}\gamma  (\pmb{g}_{k}^{n}\pmb{p}^*+\tilde{\sigma}_{n}^{2})$, then we can reduce the  power $p_{k}^*$ to
strictly improve the objective function without violating all constraints.
According to \cite[Lemma 4 (2)]{Hanly1995}, the solution to the fixed point equation  (\ref{min power_SISO_6}) is unique.
Thus $\bm p^* $ must coincide with $\hat{\bm p}$ and $ \textrm{(P\textsubscript{SISO-QoS-1})}$ has a unique optimal solution $\hat{\bm p}$.

Next, we show that $\hat{\bm p}$ is also the unique optimal solution to
 \textrm{(P\textsubscript{SISO-QoS-2})}. Assume the contrary, that \textrm{(P\textsubscript{SISO-QoS-2})} has an optimal solution $\tilde{\bm p} \neq \hat{\bm p}$. Due to the optimality of $\tilde{\bm p}$, we have
 \begin{equation}\label{tilde greater than hat p}
   \sum_k \tilde{p}_k  \geq \sum_k \hat{p}_k.
 \end{equation}
 Define a set $\mathcal{K} = \{ k \mid  \tilde{p}_k > \hat{p}_k \}$.
 According to \eqref{tilde greater than hat p} and the assumption  $\tilde{\bm p} \neq \hat{\bm p}$, the set $\mathcal{K}$ must be nonempty;
 otherwise, we have $\tilde{p}_k \leq \hat{p}_k, \forall k$, which together with \eqref{tilde greater than hat p} implies $\tilde{p}_k = \hat{p}_k$,
 a contradiction.

Define $ k_0 = \arg \max_k \left\{  \frac{\tilde{p}_k }{ \hat{p}_k } \right\}$ and
$\tau = \max_k \left\{  \frac{\tilde{p}_k }{ \hat{p}_k } \right\} = \frac{\tilde{p}_{k_0} }{ \hat{p}_{k_0} } > 1$, then
\begin{equation}\label{tau hat p larger}
  \tau \hat{p}_j \geq \tilde{p}_j,  \ \forall j.
\end{equation}
We have
 \begin{align}
 & \min_n \gamma(\bm g_{k_0}^n \tilde{\bm p} + \tilde{\sigma}_n^2 )
   \overset{(i)}{ \geq} \tilde{p}_{k_0}
   = \tau \hat{p}_{k_0}
   \overset{(ii)}{=} \tau \gamma \min_n (\bm g_{k_0}^n \hat{\bm p} + \tilde{\sigma}_n^2 ) \nonumber \\
  &   \overset{(iii)} {>}  \gamma \min_n (\bm g_{k_0}^n \tau \hat{\bm p} +  \tilde{\sigma}_n^2 )
   \overset{(iv)}{\geq} \gamma  \min_n (\bm g_{k_0}^n \tilde{\bm p} +  \tilde{\sigma}_n^2 )\nonumber .
 \end{align}
Here, $(i)$ is because $\bm \tilde{p}$ satisfies the constraints of  \textrm{(P\textsubscript{SISO-QoS-2})},
$(ii)$ is because $\hat{\bm p}$ is the fixed point of \eqref{min power_SISO_6},
$(iii)$ is due to the facts that $\tau > 1$ and the noise variance $\tilde{\sigma}_n^2 >0, \forall n $, and
$(iv)$ follows from \eqref{tau hat p larger}.
The above relation is a contradiction,  thus $\hat{\bm p}$ must be the unique optimal solution to
 \textrm{(P\textsubscript{SISO-QoS-2})}. $ \Box $

 Lemma \ref{lemma of equivalence of SISO1 and SISO2} implies that solving (P\textsubscript{SISO-QoS-2}) either provides an optimal solution to (P\textsubscript{SISO-QoS-1})  or provides an infeasibility certificate for (P\textsubscript{SISO-QoS-1}); in fact, (P\textsubscript{SISO-QoS-1}) is infeasible if and only if for any optimal solution to (P\textsubscript{SISO-QoS-2}) there is at least one active inequality. Therefore,
 Lemma \ref{lemma of equivalence of SISO1 and SISO2} leads to a two-step algorithm for solving (P\textsubscript{SISO-QoS-1}):

Step 1: Find one optimal solution $\tilde{\bm p}$ to (P\textsubscript{SISO-QoS-2}).

Step 2: Equality test: test whether $ \tilde{p}_k = \min_n \gamma (\bm g_k^n \tilde{\bm p} + \tilde{\sigma}^2_n ), \forall k $.
  If yes, then $\tilde{\bm p}$ is the unique optimal solution to  (P\textsubscript{SISO-QoS-1});
  if no, then (P\textsubscript{SISO-QoS-1}) is infeasible.

Problem (P\textsubscript{SISO-QoS-2}) can be recast  as
 \begin{equation}\label{min power_SISO_7}
\begin{split}
\textrm{(P\textsubscript{SISO-QoS-LP})}:\max_{\pmb{p} }  & \sum_{k=1}^K p_k, \\
\textrm{s.t.} \quad  & 0 \leq p_k \leq \bar{p}_k, \; k=1,\dots,K, \\
            &  p_{k} \leq \gamma  (\pmb{g}_{k}^{n}\pmb{p}+\tilde{\sigma}_{n}^{2}), \; \\
            & k=1,\dots,K,n=1,\dots,N,
\end{split}
\end{equation}
which is an LP (linear programming) and thus polynomial time solvable.
As a result, (P\textsubscript{SISO-QoS-1}) can be solved by a two-step algorithm where the first step consists of solving an LP, and the second
step is a simple equality test. Thus (P\textsubscript{SISO-QoS-1}) can be solved in polynomial time.

Let us come back to the proof of Theorem 1.
  (P\textsubscript{SISO})  can be solved by a binary search method whereby each subproblem (P\textsubscript{SISO-QoS-1})
  can be solved by an LP plus an equality test (we refer to this method as BS-LP algorithm), thus (P\textsubscript{SISO})  is polynomial time solvable.
 \end{IEEEproof}
\subsection{A Fixed Point Based Binary Search Algorithm}\label{sec:System Overview_MIMO_OSTBC_Optimal_Relay}
In the BS-LP algorithm, we need to solve a series of LPs, which may still be computationally intensive.
In this section, we present a BS-FP algorithm, which solves the QoS constraint subproblem  (P\textsubscript{SISO-QoS}) using an existing fixed point method without resorting to LPs.  To this end, define
\begin{equation}\label{optimal a1}
 T_k^{(n)}(\bm p) \triangleq \left\{ \frac{   \sigma_{n}^2 + \sum_{j \neq k} g_{n j} p_j }{ g_{n k} }  \right\} ,
 \end{equation}
 \begin{equation}\label{def of T_k}
T_k(\bm p) \triangleq \min_{ 1\leq n\leq N} T_k^{(n)}(\bm p),
 \end{equation}
\begin{equation}\label{optimal a}
 A_k (\bm p) \triangleq \arg \min_{ n } T_k^{(n)}(\bm p).
 \end{equation}
 Notice that $T_k^{(n)}(\bm p)$ represents the minimum power needed by user $k$ to achieve a SINR value of $1$ if its associated BS is $n$ and the power of other users are fixed at $p_j, \forall j\neq k$.
The minimum power user $k$ needs to achieve a SINR level of $1$ among all possible choices of BS association is defined as $ T_k(\bm p)$, and the corresponding
BS association is defined as $A_k (\bm p)$ (if there are multiple elements in $\arg \min_{ n } T_k^{(n)}(\bm p)$, let $A_k (\bm p)$  be any one of them). Note that the BS association $a_k$ defined in ($\ref{optimal a, formula}$) is precisely $A_k(\bm p)$.

Reference \cite{Yates1997} proposed a general algorithmic framework based on the standard interference functions, and we will use the fact that $T_{k}(\pmb{p})$ is a standard interference function (for completeness, see Lemma \ref{lemma of standard_interfernece_function_simo} for a proof) to apply the framework to the QoS constrained problem \textrm{(P\textsubscript{SISO-QoS})}.
The algorithm of \cite{Yates1997} starts from any positive vector $\bm p(0)$, and updates the power vector by
\begin{equation}\label{min power, update procedure}
 p_k(t+1) = \min \{ \gamma T_k (\bm p(t)),   \bar{p}_k \}, \; k=1,\dots, K ,
\end{equation}
where $\bm p(t) = (p_1(t),\dots, p_K(t))$ is the power vector at the $t$-th iteration. 
It has been shown in \cite[Section V.B,Corollary 1]{Yates1997} that the above procedure (\ref{min power, update procedure}) converges to $\bm q$, which is the unique fixed point of the following equation:
\begin{equation}\label{min power pro, fixed point iteration}
 q_k =  \min \{  \gamma T_k(\bm q),   \bar{p}_k \}, \; k=1,\dots, K,
\end{equation}

Let the corresponding BS association $b_k = A_k(\bm q) $, and denote $\gamma_{\rm ach}$ as the minimum SINR achieved by $(\bm q, \bm b)$, i.e.
\begin{equation}\label{min power pro, fixed point iteration_new_1}
\gamma_{\rm ach}=\min_{k}\frac{g_{b_k k} q_{k} }{\sigma_{b_k}^2 + \sum_{j\neq k}g_{b_k j}q_j  }=\min_{k}\frac{q_{k}}{T_{k}(\pmb{q})},\; k=1,\dots, K,
\end{equation}
Since $q_k \leq  \gamma T_k(\bm q),\; \forall k$, we have $\gamma_{\rm ach} \leq \gamma$.
\begin{proposition}\label{prop: feasibility check}
If $\gamma_{\rm ach} = \gamma$, then problem (P\textsubscript{SISO-QoS}) is feasible and $ (\bm q, \bm b) $ is an optimal solution; if $\gamma_{\rm ach} < \gamma$,
(P\textsubscript{SISO-QoS})  is infeasible. 
\end{proposition}
\begin{IEEEproof}[Proof of Proposition \ref{prop: feasibility check}]
See Appendix \ref{sec:proof_of_proposition_1}.
\end{IEEEproof}
Proposition \ref{prop: feasibility check} implies that the procedure (\ref{min power, update procedure}) can be used to check the feasibility of problem (P\textsubscript{SISO-QoS}).
 Combining the fixed point method  (\ref{min power, update procedure})  with a binary search method, the problem (P\textsubscript{SISO})
can be solved to global optima.
\subsection{A Normalized Fixed Point Algorithm}\label{sec:System Overview_MIMO_OSTBC_Optimal_Relay}
Both the BS-LP and BS-FP  algorithms  invoke the binary search, resulting in an intensive computational burden.
In this subsection, we propose a novel NFP (Normalized Fixed Point) algorithm, which can directly solve the joint BS association and power control problem without resorting to the binary search method.

Denote
 $$\bar{\bm p} \triangleq (\bar{p}_1,\dots, \bar{p}_K), $$
 $$T(\bm p) \triangleq (T_1(\bm p), \dots, T_K(\bm p)),$$
where $\bar{p}_k$ is the power budget of user $k$, and $T_k(\bm p)$ is defined in (\ref{def of T_k}).
Define a weighted infinity norm $\| \cdot \|_{\infty}^{\bar{ \bm p}}$ as
\begin{equation}\label{define infty norm}
\| \bm x \|^{ \bar{\bm p} }_{\infty} = \max_{1\leq k \leq K} \frac{x_k }{ \bar{p}_k }, \; \forall \;\bm x \in \mathbb{R}^K.
\end{equation}
If all users have the same power budget $\bar{p}_k = P_{\max}$, the defined norm $\| \bm x \|^{ \bar{\bm p} }_{\infty} = \| x\|_{\infty}/P_{\max} $.

The proposed algorithm is based on the following lemma, which states that the optimal power vector satisfies a fixed point equation.

\begin{lemma}\label{lemma of fixed point}
Suppose $(\bm p^*, \bm a^*)$ is an optimal solution to problem \textrm{(P\textsubscript{SISO})}, then $\bm p^*$ satisfies the following equation:
\begin{equation}\label{fixed point of max min}
\bm p^* = \frac{T(\bm p^*)}{\| \bm T(\bm p^*) \|^{ \bar{\bm p} }_{\infty}  }.
\end{equation}
\end{lemma}
\noindent\emph{Proof of Lemma \ref{lemma of fixed point}}:
For a given power allocation $\bm p^*$, the optimal BS association is $a_k^* =  A_k (\bm p^*) = \arg \min_{ n } T_k^{(n)}(\bm p^*).$
Therefore, the SINR of user $k$ at optimality is
\begin{equation}\label{expression of SINR_k*}
 \text{SINR}_k^* = \frac{p_k^* }{ T_k^{(a^*_k)}(\bm p^*) } =\frac{ p_k^* }{  \min_{ n } T_k^{(n)}(\bm p^*) } = \frac{ p_k^*  }{ T_k(\bm p^*)}.
\end{equation}

Let $ \gamma^* $ denote the optimal value $ \min_k \text{SINR}_k^* $, then we have
\begin{equation}\label{SINR_k are equal}
\text{SINR}_k^* = \gamma^*,\quad \forall\; k.
\end{equation}
In fact, if $\text{SINR}_j^* > \gamma^*$ for some $j$, then we can reduce the power of user $j$ so that $\text{SINR}_j$ decreases and all other $\text{SINR}_k$ 's increase, yielding a minimum SINR that is higher than $\gamma^*$. This contradicts the optimality of $\gamma^*$, thus (\ref{SINR_k are equal}) is proved.

According to (\ref{expression of SINR_k*}) and (\ref{SINR_k are equal}), we have
\begin{equation}\label{T over p is gamma^*}
 \gamma^* T_k(\bm p^*)   = p_k^*, \quad \forall\; k.
\end{equation}

Next, we show that at least one user transmits at full power, i.e.
\begin{equation}\label{one power tight}
\max_k \frac{ p_k^*}{ \bar{p}_k } = 1.
\end{equation}
Assume $ \mu = \max_k \frac{ p_k^*}{\bar{p}_k} < 1 $. Define a new power vector $ \bm p = \bm p^*/\mu$, then $\bm p$ satisfies the power constraints $p_k \leq \bar{p}_k, \forall k$. The SINR of user $k$ achieved by $(\bm p, \bm a^*)$ is
 $\text{SINR}_k = p_k/T_k^{ (a^*_k) }(\bm p)  =  p^*_k / ( \mu T_k^{ (a^*_k) }(\bm p^* /\mu)  ) > p^*_k / ( T_k^{ (a^*_k) }(\bm p^*) ) = \text{SINR}_k^* $,
 which contradicts the optimality of $(\bm p^*, \bm a^*)$.

Plugging (\ref{T over p is gamma^*}) into (\ref{one power tight}), we obtain
\begin{equation}\label{gamma* expressed by norm of T}
\frac{ 1 } {\gamma^* } = \max_k \frac{ T_k(\bm p^*) }{\bar{p}_k} = \|T(\bm p^*) \|_{\infty}^{\bar{p}}.
\end{equation}
Combining (\ref{T over p is gamma^*}) and (\ref{gamma* expressed by norm of T}), we obtain (\ref{fixed point of max min}).  $\blacksquare$

Based on the fixed point equation (\ref{fixed point of max min}), we propose an  NFP algorithm to solve problem (\ref{max min}).
\begin{table}[htbp]
 \caption{NFP Algorithm for UL SISO Cellular Networks}
 \label{table:table of normalized fixed point_SISO}
 \large
 \begin{tabular}{p{460pt}}
\hline
Initialization: pick random positive power vector $\bm p(0)$.
\\
Loop $t$:
\\ 1) Compute BS association: $a_k(t) \leftarrow A_k(\bm p(t)), \ \forall\; k $.
\\ 2) Update power: $ \bm p(t+1) \leftarrow  T( \bm p(t))$ ;
\\ 3) Normalize: $ \bm p(t+1) \leftarrow \frac{\bm p(t+1)}{ \| \bm p(t+1) \|_{\infty}^{\bar{p}} }  $ ,   where $ \| \bm p(t+1) \|_{\infty}^{\bar{p}} = \max_k \frac{ p_k(t+1) }{ \bar{p}_k} $.
\\ Iterate until convergence.
\\
\hline
\end{tabular}
\end{table}
The following theorem shows that the NFP algorithm in Table \ref{table:table of normalized fixed point_SISO} converges to the optimal solution to (\ref{max min}) at a geometric rate.
\begin{theorem}\label{thm of convergence}
Suppose $(\bm p^*, \bm a^*)$ is an optimal solution to problem \textrm{(P\textsubscript{SISO})}.
Then the sequence $\{\bm p(t)\} $ generated by the NFP algorithm in Table \ref{table:table of normalized fixed point_SISO}  converges geometrically to $\bm p^*$, i.e.,
\begin{equation}\label{converg speed}
 \|\bm p(t) - \bm p^* \|_{\infty}^{\bar{p}} \leq C  \kappa^t,
\end{equation}
where $C>0,\ 0< \kappa < 1$ are constants that depend only on the problem data. 
\end{theorem}

\noindent\emph{Proof of Theorem \ref{thm of convergence}}: By definition \eqref{optimal a1}, the mapping
 $T(\bm p)=(T_1(\bm p), \dots, T_K(\bm p)): \mathbb{R}_{+}^K \rightarrow \mathbb{R}_{+}^K$ is the pointwise minimum of affine linear mappings $T^{(n)}(\bm p ) = (T_1^{(n)}(\bm p ), \dots, T_K^{(n)}(\bm p) )$, for $n=1,\dots, N$. It follows that $T$ is a concave mapping. According to Lemma~\ref{lemma of fixed point}, $\bm p^*$ is a fixed point of (\ref{fixed point of max min}).
According to the concave Perron-Frobenius theory \cite[Theorem 1]{krause2001concave}, (\ref{fixed point of max min}) has a unique fixed point, and the NFP algorithm in Table \ref{table:table of normalized fixed point_SISO} converges to this fixed point. Therefore, the NFP algorithm in Table \ref{table:table of normalized fixed point_SISO}  converges to $\bm p^*$.

To show the geometric convergence, we define $U$ as the set of power vectors $\bm p$ with $\|\bm p \|_{\infty}^{\bar{p}} = 1$ (i.e. $\max_k \frac{p_k}{\bar{p}_k} = 1$).
It can be easily verified that
 \begin{equation}\label{lower upper}
  A_k \leq T_k(\bm p) \leq  B_k , \quad \forall\; \bm p \in U ,
  \end{equation}
  where $ A_k = \min_n \frac{\sigma_n^2  }{g_{nk}} $ and $B_k =  T_k( \bar{\bm p})   
  =\min_n \frac{\sigma_n^2 + \sum_{j \neq k} g_{n j } \bar{p}_j }{g_{nk}}$ are both constants that only depend on the problem data.
  For two vectors $x, y$, we denote $x \geq y $ if $x_k \geq y_k,\ \forall\; k$.
   Define $\kappa = 1-\min_k \frac{A_k}{B_k} \in (0,1)$  and $\bm e = (B_1,\dots, B_K) >0 $.
  Then (\ref{lower upper}) implies
 \begin{equation}\label{boundness of T}
  (1 - \kappa) e \leq T(\bm p) \leq e , \quad \forall\; \bm p \in U .
 \end{equation}
 According to the concave Perron-Frobenius theory  \cite[Lemma 3, Theorem]{krause1994relative},
if $T$ is a concave mapping and satisfies (\ref{boundness of T}), then the NFP algorithm in Table \ref{table:table of normalized fixed point_SISO}  converges geometrically at the rate $\kappa$.  $\blacksquare$

\begin{remark}\label{pseudo_polynomial_time_siso}
 Theorem \ref{thm of convergence} implies the pseudo-polynomial time solvability of problem \eqref{max min}.
Without loss of generality, we can assume $\sigma_n^2 = 1$; in fact, replacing $g_{nk}^2$ by $g_{nk}^2/\sigma_n^2$ and $\sigma_n^2$ by $1$ for all $n,k$ does not change problem \eqref{max min} and
the NFP algorithm in Table \ref{table:table of normalized fixed point_SISO}.
It is easy to verify that $ \kappa \leq 1 - 1/( K G \cdot \mathrm{SNR} +1 ) $, where $\mathrm{SNR} = \max_{k} \bar{p}_k$ and $G = \max_{n,k} \{ g_{nk} \} $. 
 To achieve an $\epsilon$-optimal solution, the NFP algorithm in Table \ref{table:table of normalized fixed point_SISO} takes
 $T \leq \frac{\log(1/\epsilon)}{ \log( 1/\kappa ) } \leq \log(1/\epsilon)( K G \cdot \mathrm{SNR}+1)$ iterations.
 Since $K G \cdot \mathrm{SNR}$ is polynomial in the input parameters $K, \{\bar{p}_k\}$ and $\{ g_{nk}\} $, we obtain the pseudo-polynomial time solvability of problem \eqref{max min}.  Note that to prove the polynomial time solvability, we need to show that $T$ is upper bounded by a polynomial function of
 $K, \{\log p_{k}\},\{\log g_{nk}\}$.   It is an open question whether the NFP algorithm is a polynomial time algorithm or not, though we observe that the NFP algorithm always converges much faster than the polynomial time algorithm BS-LP in the numerical experiments.

 \end{remark}
\section{Joint BS Association and Beamforming for UL SIMO Cellular Networks}\label{sec:Joint_BS_Association_and_Beamforming_for_SIMO_System}
For a SIMO system where  $M_{n}>1,\ \forall \ n$, the beamforming vectors $\{\pmb{u}_{n,k}\}$ are also design variables, making the problem (P) much more complicated than the SISO scenario.
For fixed $\{\pmb{u}_{n,k}\}$, the problem reduces to a joint BS association and power control design problem, which can be solved by the algorithms dedicated
to the SISO scenario.
 For a fixed power $\pmb{p}$, the  optimal receiver beamforming vector ${\pmb{u}}_{n,k}$ is given by  \cite{Ya-Feng-Liu-1}
 \begin{align}\label{eq:uplink_assocation_miso_sysyem_model_5}
\hat{\pmb{u}}_{n,k}=\pmb{M}_{n}^{-1}(\pmb{p})\pmb{h}_{n,k}
\end{align}
up to a scaling factor (note that the optimal $\pmb{u}_{n,k}$ is independent of $\pmb{a}$), where $\pmb{M}_{n}(\pmb{p})$ is given by \cite{Ya-Feng-Liu-1}
 \begin{align}\label{eq:uplink_assocation_miso_sysyem_model_5-2}
\pmb{M}_{n}(\pmb{p})=\sigma_{n}^2\pmb{I}+\sum_{j=1}^{K}\pmb{h}_{n,j}\pmb{h}_{n,j}^{H}p_{j},
\end{align}
 and the optimal association vector $\pmb{a}$ is given by
\begin{align}\label{eq:uplink_assocation_miso_sysyem_model_6}
a_{k}=\textrm{arg} \max_{n \in \{1,\dots, N\}} \left\{\frac{p_{k}\hat{\pmb{u}}_{n,k}^{H}\pmb{h}_{n,k}\pmb{h}_{n,k}^{H}\hat{\pmb{u}}_{n,k}}{\sigma_{n}^{2}+\hat{\pmb{u}}_{n,k}^{H}\sum_{j=1,j\not=k}^{K}p_{j}\pmb{h}_{n,j}\pmb{h}_{n,j}^{H}\hat{\pmb{u}}_{n,k}}\right\}.
\end{align}
 For fixed association profile $\pmb{a}$,  the problem reduces to maximizing the minimum $\textrm{SINR}$
  by jointly designing $\pmb{p}$ and  receiver beamforming vectors $\{\pmb{u}_{a_{k},k}\}$, which is polynomial time solvable \cite{Ya-Feng-Liu-1}.
Specifically,  the optimal receiver beamforming vectors  $\{\hat{\pmb{u}}_{a_{k},k}\}$ can be given by \cite{Ya-Feng-Liu-1}
  \begin{align}\label{eq:uplink_assocation_miso_sysyem_model_5-1}
\hat{\pmb{u}}_{a_{k},k}=\pmb{M}_{a_{k}}^{-1}(\pmb{p})\pmb{h}_{a_{k},k}
\end{align}
up to a scaling factor.  Upon substituting (\ref{eq:uplink_assocation_miso_sysyem_model_5-1})   into
(\ref{eq:uplink_assocation_miso_sysyem_model_3}),  the SINR for the $k$th user can be expressed as
\begin{align}\label{eq:uplink_assocation_miso_sysyem_model_5-2}
\textrm{SINR}_{k}=\frac{1}{\frac{1}{p_{k}\pmb{h}_{a_{k},k}^{H}\pmb{M}_{a_{k}}^{-1}(\pmb{p})\pmb{h}_{a_{k},k}}-1}.
\end{align}
The problem becomes maximizing the minimum $\textrm{SINR}$ of (\ref{eq:uplink_assocation_miso_sysyem_model_5-2}) over $\pmb{p}$, which can be solved by a BS-SDP algorithm in \cite{Ya-Feng-Liu-1}.

However, when joint designing BS association, power control and beamforming vectors, the problem  (P) becomes  more complicated, which is the focus of this section.
\subsection{Polynomial Time Solvability}\label{sec:System Overview_MIMO_OSTBC_Optimal_Relay}
In this section, we will prove that the problem (P) is polynomial solvable for the SIMO scenario.  While the overall proof framework is similar to the proof of Theorem \ref{thm_polynomial_time_solvable},
for Theorem \ref{thm_polynomial_time_solvable_simo} we need to deal with the extra operator $\pmb{M}_k^{-1}(\pmb{p})$  which does not have closed form. This makes the proof more involved than the proof of Theorem  \ref{thm_polynomial_time_solvable}. For example, the uniqueness of the solution in Lemma \ref{lemma of polynomial_time_2} was not known before, and we utilize the property of $\pmb{M}_k^{-1}(\pmb{p})$ to derive a new proof.
\begin{theorem}\label{thm_polynomial_time_solvable_simo}
Problem (P), i.e., the maximizing the minimum SINR by joint BS association, power control and beamforming for an uplink SIMO cellular network, is polynomial time solvable.
\end{theorem}
\begin{IEEEproof}[Proof of Theorem \ref{thm_polynomial_time_solvable_simo}]
Let us consider the following power minimization problem with QoS constraints:
\begin{subequations}
\begin{align}
&\textrm{(P\textsubscript{SIMO-QoS})}: \min_{\pmb{p},\pmb{a},\{\pmb{u}_{n,k}\}_{k=1,\dots,K, \; n=1,\dots,N} }   \sum_{k=1}^K p_k,\nonumber \\
&\textrm{s.t.} \quad   0 \leq p_k \leq \bar{p}_k, \; k=1,\dots,K, \label{eq:subeq1_qos} \\
&                a_k \in \{1,2,\dots, N \}, \; k=1,\dots,K,\label{eq:subeq2_qos} \\
&           \textrm{SINR}_{k}=\frac{p_{k}\pmb{u}_{a_{k},k}^{H}\pmb{h}_{a_{k},k}\pmb{h}_{a_{k},k}^{H}\pmb{u}_{a_{k},k}}{\sigma_{a_{k}}^{2}+\pmb{u}_{a_{k},k}^{H}\sum_{j=1,j\not=k}^{K}p_{j}\pmb{h}_{a_{k},j}\pmb{h}_{a_{k},j}^{H}\pmb{u}_{a_{k},k}}\geq\gamma, \;\label{eq:subeq3_qos} \\
&           k=1,\dots, K,\nonumber\\
&             \|\pmb{u}_{n,k}\|=1, \; k=1,\dots,K,\; n=1,\dots,N, \label{eq:subeq4-2_qos}
\end{align}
\end{subequations}
where  $\gamma$ is the required SINR value.
Similar as the SISO scenario,  Problem (P) can be solved  by a sequence of subproblems of the form \textrm{(P\textsubscript{SIMO-QoS})} and a binary search on $\gamma$.   In the following, we will show that the QoS constrained subproblem \textrm{(P\textsubscript{SIMO-QoS})} can be transformed to a semidefinite  programming (SDP).

For fixed power $\pmb{p}$, based on the expression of  the optimal receiver beamforming vector $\hat{\pmb{u}}_{n,k}$ in (\ref{eq:uplink_assocation_miso_sysyem_model_5}), we have
\begin{align}
&\exists \ \{\pmb{u}_{n,k}\}, \textrm{satisfying}~  \textrm{ (\ref{eq:subeq3_qos}) and  (\ref{eq:subeq4-2_qos})}\nonumber \\
&\Longleftrightarrow
 \frac{p_{k}\pmb{h}_{a_{k},k}^{H}\pmb{M}_{a_{k}}^{-1}(\pmb{p})\pmb{h}_{a_{k},k}}{1-p_{k}\pmb{h}_{a_{k},k}^{H}\pmb{M}_{a_{k}}^{-1}(\pmb{p})\pmb{h}_{a_{k},k}} \geq \gamma, \; k=1,\dots,K\\ &\Longleftrightarrow
  p_{k} \pmb{h}_{a_{k},k}^{H}\pmb{M}_{a_{k}}^{-1}(\pmb{p})\pmb{h}_{a_{k},k} \geq \frac{\gamma}{1+\gamma}, \; k=1,\dots,K. \label{eq:subeq4-11_qos}
\end{align}
Since the optimal BS association $a_{k}$ is given by (\ref{eq:uplink_assocation_miso_sysyem_model_6}), we have
\begin{align}
& \exists \ \pmb{a}~~\textrm{satisfying}~~\textrm{(\ref{eq:subeq2_qos}) and (\ref{eq:subeq4-11_qos})} \nonumber\\
&\Longleftrightarrow
\max_{n \in \{1,\dots, N\}}  p_{k} \pmb{h}_{n,k}^{H}\pmb{M}_{n}^{-1}(\pmb{p})\pmb{h}_{n,k}   \geq \frac{\gamma}{1+\gamma},\; k=1,\dots,K
\end{align}
Consequently,  the problem (P\textsubscript{SIMO-QoS}) is equivalent to the following problem:
 \begin{equation}\label{eq:subeq4_qos}
\begin{split}
& \textrm{(P\textsubscript{SIMO-QoS-1})}:\min_{\pmb{p} }   \sum_{k=1}^K p_k, \\
\textrm{s.t.} \quad  & 0 \leq p_k \leq \bar{p}_k, \; k=1,\dots,K, \\
            & \max_{n \in \{1,\dots, N\}}  p_{k} \pmb{h}_{n,k}^{H}\pmb{M}_{n}^{-1}(\pmb{p})\pmb{h}_{n,k}   \geq \frac{\gamma}{1+\gamma}, \; \\
             & k=1,\dots,K.
\end{split}
\end{equation}
\begin{lemma}\label{lemma of polynomial_time_2-1}
The equation
 \begin{equation}\label{eq:subeq4-1_qos}
\max_{n \in \{1,\dots, N\}}  p_{k} \pmb{h}_{n,k}^{H}\pmb{M}_{n}^{-1}(\pmb{p})\pmb{h}_{n,k}   = \frac{\gamma}{1+\gamma}, \; k=1,\dots,K.
\end{equation}
has a unique fixed point.
\end{lemma}
\begin{IEEEproof}[Proof of Lemma \ref{lemma of polynomial_time_2-1}]
See Appendix \ref{sec:proof_of_lemma_3-1}.
\end{IEEEproof}

\begin{proposition}\label{prop: feasibility check_simo-0}
For a given $\gamma$, if problem (P\textsubscript{SIMO-QoS-1}) is feasible, its optimal solution $\hat{\pmb{p}}$ must satisfy (\ref{eq:subeq4-1_qos}), i.e.,
\begin{align}\label{eq:subeq6-1_qos}
\max_{n}  \hat{p}_{k} \pmb{h}_{n,k}^{H}\pmb{M}_{n}^{-1}(\hat{\pmb{p}})\pmb{h}_{n,k} =\frac{\gamma}{1+\gamma} , \ k=1,\dots,K,
\end{align}
and $\hat{\pmb{p}}$ is the unique solution to problem (P\textsubscript{SIMO-QoS-1}).
\end{proposition}
\begin{IEEEproof}[Proof of Proposition \ref{prop: feasibility check_simo-0}]
If (\ref{eq:subeq6-1_qos}) does not hold for some $k$, according to \cite[Lemma 3.1]{Ya-Feng-Liu-1},
we can reduce $\hat{p}_{k}$ to achieve a lower objective value without violating any constraints.  According to Lemma \ref{lemma of polynomial_time_2-1},  such $\hat{\pmb{p}}$ is unique.
\end{IEEEproof}

Reversing the direction of the inequality in the second constraint above and maximizing the objective function instead of minimizing, we obtain a new problem
 \begin{equation}\label{eq:subeq5_qos}
\begin{split}
& \textrm{(P\textsubscript{SIMO-QoS-2})}:\max_{\pmb{p} }   \sum_{k=1}^K p_k,\\
\textrm{s.t.} \quad  & 0 \leq p_k \leq \bar{p}_k, \; k=1,\dots,K, \\
            & \max_{n \in \{1,\dots, N\}}  p_{k} \pmb{h}_{n,k}^{H}\pmb{M}_{n}^{-1}(\pmb{p})\pmb{h}_{n,k}   \leq \frac{\gamma}{1+\gamma}, \;\\
            & k=1,\dots,K,
\end{split}
\end{equation}
which is always feasible since $(0,0,\dots,0)$ is one feasible solution.
The following result shows that   \textrm{(P\textsubscript{SIMO-QoS-1})} and \textrm{(P\textsubscript{SIMO-QoS-2})} are ``equivalent''.

\begin{lemma}\label{lemma of polynomial_time_2}
The two problems \textrm{(P\textsubscript{SIMO-QoS-1})} and \textrm{(P\textsubscript{SIMO-QoS-2})} are equivalent in the following sense:
for a given $\gamma$, if \textrm{(P\textsubscript{SIMO-QoS-1})} is infeasible, then for any optimal solution to \textrm{(P\textsubscript{SIMO-QoS-2})},
denoted as $\tilde{\bm p}$, there exists some $k$ such that $ \max_{n \in \{1,\dots, N\}}  \tilde{p}_{k} \pmb{h}_{n,k}^{H}\pmb{M}_{n}^{-1}(\tilde{\pmb{p}})\pmb{h}_{n,k}   < \frac{\gamma}{1+\gamma} $;
if problems (P\textsubscript{SIMO-QoS-1}) is feasible, then $\hat{\bm p} $  is the unique optimal solution to both \textrm{(P\textsubscript{SIMO-QoS-1})}
and  \textrm{(P\textsubscript{SIMO-QoS-2})}, where $ \hat{\bm p} $ is the unique fixed point of (\ref{eq:subeq4-1_qos}).
\end{lemma}
\begin{IEEEproof}[Proof of Lemma \ref{lemma of polynomial_time_2}]
See Appendix \ref{sec:proof_of_lemma_3}.
\end{IEEEproof}
 Problem (P\textsubscript{SIMO-QoS-2}) can be rewritten  as
 \begin{equation}\label{eq:subeq11_qos}
\begin{split}
\max_{\pmb{p} }  & \sum_{k=1}^K p_k,\\
\textrm{s.t.} \quad  & 0 \leq p_k \leq \bar{p}_k, \; k=1,\dots,K, \\
            & p_{k}\pmb{h}_{n,k}^{H}\pmb{M}_{n}^{-1}(\pmb{p})\pmb{h}_{n,k} \leq \frac{\gamma}{1+\gamma}, \;\\
            &  k=1,\dots,K,\ n=1,\dots,N.
\end{split}
\end{equation}
By using Schur complement,  the above problem  can be further rewritten  in the SDP format as (note that (\ref{eq:subeq12_qos}) is an SDP since $\pmb{M}_n(\pmb{p})$ defined by (\ref{eq:uplink_assocation_miso_sysyem_model_5-2}) depends linearly on $\pmb{p}$)  \cite{Ya-Feng-Liu-1}
 \begin{equation}\label{eq:subeq12_qos}
\begin{split}
\textrm{(P\textsubscript{SIMO-QoS-SDP})}:\max_{\pmb{p} }  & \sum_{k=1}^K p_k,\\
\textrm{s.t.} \quad  & 0 \leq p_k \leq \bar{p}_k, \; k=1,\dots,K, \\
            &  \pmb{M}_{n}(\pmb{p}) \succeq p_{k}(1+\frac{1}{\gamma})\pmb{h}_{n,k}\pmb{h}_{n,k}^{H},    \; \\
            & k=1,\dots,K, \ n=1,\dots,N,
\end{split}
\end{equation}
which is polynomial time solvable.

 Lemma \ref{lemma of polynomial_time_2} and Proposition \ref{prop: feasibility check_simo-0}  suggest a two-step algorithm to solve (P\textsubscript{SIMO-QoS-1}) \cite{Ya-Feng-Liu-1}:

Step 1: SDP: For a given $\gamma$, solve the problem (P\textsubscript{SIMO-QoS-SDP}) and denote its optimal solution as  $\hat{\pmb{p}}$.

Step 2: Equality test: test whether $\hat{\pmb{p}}$ satisfies (\ref{eq:subeq6-1_qos}).
  If yes, then $\hat{\bm p}$ is the unique optimal solution to  (P\textsubscript{SISO-QoS-1});
  if no, then (P\textsubscript{SIMO-QoS-1}) is infeasible.

  Let us come back to the proof of Theorem 3.   Problem (P) can be solved by a binary search method whereby each subproblem (P\textsubscript{SIMO-QoS-1}) can be solved by an SDP and equality test (we refer to this method as BS-SDP algorithm), thus (P) is also  polynomial time solvable.
\end{IEEEproof}
\subsection{A Fixed Point Based Binary Search Algorithm}\label{sec:System Overview_MIMO_OSTBC_Optimal_Relay}
In the BS-SDP algorithm, solving a series of SDPs  may impose an intensive computational burden.  In this section, we present a  BS-FP algorithm, which solves the QoS constrained subproblem (P\textsubscript{SIMO-QoS}) using a fixed point method without invoking SDP.   This algorithm is a direct generalization of the BS-FP algorithm for the SISO case. However, this algorithm was not explicitly stated in the literature; in addition, we are not aware of an explicit statement and proof of Lemma \ref{lemma of standard_interfernece_function_simo} in previous works, even though it is probably not surprising for experts in this area. To this end,
denote
\begin{align}\label{eq:uplink_assocation_miso_minimum_power_model_5}
\tilde{T}_{k}^{n}(\pmb{p})& \triangleq \min_{\|\pmb{u}_{n,k}\|=1} \left\{    \frac{\sigma_{n}^{2}+\pmb{u}_{n,k}^{H}\sum_{j=1,j\not=k}^{K}p_{j}\pmb{h}_{n,j}\pmb{h}_{n,j}^{H}\pmb{u}_{n,k}}{\pmb{u}_{n,k}^{H}\pmb{h}_{n,k}\pmb{h}_{n,k}^{H}\pmb{u}_{n,k}}\right\}\nonumber\\
&=   \frac{\sigma_{n}^{2}+\hat{\pmb{u}}_{n,k}^{H}\sum_{j=1,j\not=k}^{K}p_{j}\pmb{h}_{n,j}\pmb{h}_{n,j}^{H}\hat{\pmb{u}}_{n,k}}{\hat{\pmb{u}}_{n,k}^{H}\pmb{h}_{n,k}\pmb{h}_{n,k}^{H}\hat{\pmb{u}}_{n,k}}, \end{align}
which represents the minimum power needed by user $k$ to achieve a SINR value of $1$ if its associated BS is $n$ and the power of other users are fixed at
$p_{j},\ \forall \ j\not=k$ and the optimal receiver beamforming vector $\hat{\pmb{u}}_{n,k}$ is determined by (\ref{eq:uplink_assocation_miso_sysyem_model_5}).
\begin{lemma}\label{lemma of standard_interfernece_function_simo}
$\tilde{T}_{k}^{n}(\pmb{p})$ is a standard interference function (see the definition in  \cite[Definition]{Yates1997}).
\end{lemma}

\begin{IEEEproof}
See Appendix \ref{sec:Appendix_canonical form of_MIMO_OSTBC_2}.
\end{IEEEproof}

Denote
\begin{align}\label{eq:uplink_assocation_miso_minimum_power_model_7}
\tilde{T}_{k}(\pmb{p})\triangleq \min_{1 \leq n \leq N} \tilde{T}_{k}^{n}(\pmb{p}),
\end{align}
\begin{align}\label{eq:uplink_assocation_miso_minimum_power_model_9}
\tilde{A}_{k}(\pmb{p})\triangleq \textrm{arg} \min_{n} \tilde{T}_{k}^{n}(\pmb{p}),
\end{align}
where  $\tilde{T}_{k}(\pmb{p})$ represents the minimum power user $k$ needs to achieve SINR level of $1$ among all possible choices of BS association, and the
corresponding BS association is defined as $\tilde{A}_{k}(\pmb{p})$ (if there are multiple elements in $\textrm{arg} \min_{n} \tilde{T}_{k}^{n}(\pmb{p})$, let $\tilde{A}_{k}(\pmb{p})$  be any one of them).

Since $\tilde{T}_{k}^{n}(\pmb{p})$ is a standard interference function, $\tilde{T}_{k}(\pmb{p})$ is a standard interference
function as well \cite[Theorem 5]{Yates1997}.   We apply the algorithmic framework of \cite{Yates1997} to propose the following algorithm:  starting from any positive vector $\pmb{p}(0)$,
update the power vector $\pmb{p}$  as
\begin{align}\label{eq:uplink_assocation_miso_minimum_power_model_10}
p_k(t+1)=\min \{\gamma\tilde{T}_{k}(\pmb{p}(t)), \bar{p}_{k}\},
\end{align}
where $\pmb{p}(t)=[p_{1}(t),\cdots,p_{K}(t)]$ denotes the power vector at the $t$-th iteration.
According to \cite[Section V.B, Corollary 1]{Yates95}, the algorithm (42) converges to a unique fixed point $\pmb{q}$,
 which is the unique
fixed point for
\begin{align}\label{eq:uplink_assocation_miso_minimum_power_model_11}
q_k=\min \{\gamma\tilde{T}_{k}(\pmb{p}), \bar{p}_{k}\},\ k=1,\cdots, K.
\end{align}
Denote $\pmb{b}=[b_{1}, \cdots, b_{K}]$ as the association profile corresponding to $\pmb{q}$, where $b_{k}=\tilde{A}_{k}(\pmb{p}^{*})$, and denote
$\gamma_{ach}$ as the minimum SINR achieved by  $(\pmb{q}, \pmb{b})$.
\begin{proposition}\label{prop: feasibility check_simo}
If $\gamma_{\rm ach} = \gamma$, then problem (P\textsubscript{SIMO-QoS}) is feasible and $ (\bm q, \bm b) $ is an optimal solution; if $\gamma_{\rm ach} < \gamma$,
(P\textsubscript{SIMO-QoS})  is infeasible. 
\end{proposition}
The proof of Proposition \ref{prop: feasibility check_simo} is similar to that for Proposition \ref{prop: feasibility check}.
Consequently, combining (\ref{eq:uplink_assocation_miso_minimum_power_model_10}) with a binary search method,
 problem (P)  can be solved to global optima.
\subsection{A Normalized Fixed Point Algorithm}\label{sec:System Overview_MIMO_OSTBC_Optimal_Relay}
In both the BS-SDP and BS-FP algorithms, the binary search could require significant computational burden.  In this subsection
 we propose an  NFP algorithm, which can directly solve the joint BS association, power control and beamforming problem without resorting to the binary search.
Again, this algorithm is a generalization of NFP algorithm for the SISO case.
The most nontrivial part is the proof of Lemma \ref{lemma of fixed point_simo} stated later, which is based on a technical result proved recently in \cite{Ya-Feng-Liu-1}.
With Lemma \ref{lemma of fixed point_simo}, the proof of the main result in this subsection Theorem  \ref{thm of convergence_simo} is a rather direct extension of Theorem \ref{thm of convergence}.
Define
\begin{align}\label{eq:A Normalized Fixed Point Iteration-2}
\tilde{T}(\pmb{p})\triangleq(\tilde{T}_{1}(\pmb{p}),\tilde{T}_{2}(\pmb{p}),\dots, \tilde{T}_{K}(\pmb{p})).
\end{align}
\begin{lemma}\label{lemma of fixed point_simo}
Suppose $(\pmb{p}^{*}, \{\pmb{u}_{n,k}^{*}\}_{k=1,\dots,K,\ n=1,\dots, N}, \pmb{a}^{*})$ is an optimal solution to problem (P), then $\pmb{p}^{*}$ satisfies the following equation:
\begin{align}\label{eq:A Normalized Fixed Point Iteration-4}
\pmb{p}^{*}=\frac{\tilde{T}(\pmb{p}^{*})}{\|\tilde{T}(\pmb{p}^{*})\|_{\infty}^{\bar{\pmb{p}}}}.
\end{align}
\end{lemma}
\begin{IEEEproof}[Proof of Lemma \ref{lemma of fixed point_simo}]
 For a given power allocation $\pmb{p}^{*}$, the optimal BS association $a_{k}^{*}=
\tilde{A}_{k}(\pmb{p}^{*})=\textrm{arg} \min_{n} \tilde{T}_{k}^{n}(\pmb{p}^{*})$.
Let $\textrm {SINR}_{k}^{*}$ denote the SINR of the user $k$ at the optimality.  For optimal association profile $\pmb{a}^{*}$ and optimal beamforming vector $\pmb{u}_{a_{k}^{*},k}$, according to (\ref{eq:uplink_assocation_miso_sysyem_model_5-2}),  we have
\begin{align}\label{eq:A Normalized Fixed Point Iteration-5}
\textrm {SINR}_{k}^{*}=\frac{1}{\frac{1}{p_{k}^{*}\pmb{h}_{a_{k}^{*},k}^{H}\pmb{M}_{a_{k}^{*}}^{-1}(\pmb{p}^{*})\pmb{h}_{a_{k}^{*},k}}-1}.
\end{align}

Let $\gamma^{*}$ denote the optimal value $\min_{k}\textrm {SINR}_{k}^{*}$, then we can prove
\begin{align}\label{eq:A Normalized Fixed Point Iteration-6}
\textrm {SINR}_{k}^{*}=\gamma^{*}, \ \forall \ k.
\end{align}
In fact, if $\textrm {SINR}_{j}^{*}>\gamma^{*}$, we can reduce the power of user $j$, so that $\textrm {SINR}_{j}^{*}$ decreases while the other $\textrm {SINR}_{k}^{*}$'s increase, resulting in a
minimum $\textrm{SINR}_{k}^{*}$ that is higher than $\gamma^{*}$.  Here, we use the fact that $\textrm {SINR}_{j}^{*}$ is  a strictly increasing function in $p_{j}\geq 0$ and $\textrm {SINR}_{k}^{*}$ is a strictly decreasing function in $p_{j}\geq 0, \ \forall \ k \not= j$ \cite[Lemma 3.1]{Ya-Feng-Liu-1}
(The original version of \cite[Lemma 3.1]{Ya-Feng-Liu-1} only claims that $\textrm {SINR}_{k}^{*}$  is a decreasing function on $p_{j}\geq 0,\ \forall \ k \not= j$.
However, when the entries in $\pmb{h}_{n,j}$ are  generic (e.g., drawn from a continuous probability distribution),
  $\sum_{j=1}^{K}\pmb{h}_{n,j}\pmb{h}_{n,j}^{H}p_{j}$ is a positive definite matrix  with probability one, in which case we can prove that $\textrm {SINR}_{k}$  is a strictly decreasing function on $p_{j}\geq 0,\ \forall \ k \not= j$).

Note that $\textrm {SINR}_{k}^{*}$ can also be expressed as
\begin{align}\label{eq:A Normalized Fixed Point Iteration-6-1}
\textrm {SINR}_{k}^{*}=\frac{p_{k}^{*}}{\tilde{T}_{k}^{a_{k}^{*}}(\pmb{p}^{*})}.
\end{align}
According to (\ref{eq:A Normalized Fixed Point Iteration-6}) and (\ref{eq:A Normalized Fixed Point Iteration-6-1}), we have
\begin{align}\label{eq:A Normalized Fixed Point Iteration-7}
\gamma^{*} \tilde{T}_{k}(\pmb{p}^{*})= p_{k}^{*}, \ \forall \ k.
\end{align}
Next, we show that at least one user transmits at full power, i.e.,
\begin{align}\label{eq:A Normalized Fixed Point Iteration-8}
\max_{k}\frac{p_{k}^{*}}{\bar{p}_{k}}=1.
\end{align}
Assume $\mu=\max_{k}\frac{p_{k}^{*}}{\bar{p}_{k}}<1$.  Define a new power $\pmb{p}=\frac{\pmb{p}^{*}}{\mu}$, then $\pmb{p}$ satisfies the power constraints
$p_{k}\leq \bar{p}_{k},\ \forall \ k$.
For  given $\pmb{a}^{*}$ and $\pmb{p}$, the SINR of the user $k$ achieved  can be expressed as
\begin{align}\label{eq:A Normalized Fixed Point Iteration-9}
\textrm {SINR}_{k}=\frac{p_{k}}{\tilde{T}_{k}^{a_{k}^{*}}(\pmb{p})}=\frac{p_{k}^{*}}{\mu \tilde{T}_{k}^{a_{k}^{*}}(\pmb{p}^{*}/\mu)}>\frac{p_{k}^{*}}{\tilde{T}_{k}^{a_{k}^{*}}(\pmb{p}^{*})}=\textrm {SINR}_{k}^{*},
\end{align}
where the last inequality is due to (\ref{eq:uplink_assocation_miso_minimum_power_model_6}) proved in the appendix. The above relation contradicts the optimality of (\ref{eq:A Normalized Fixed Point Iteration-9}), thus the assumption $\max_{k}\frac{p_{k}^{*}}{\bar{p}_{k}}<1$ does not hold. Therefore,
we have proved (\ref{eq:A Normalized Fixed Point Iteration-8}).

Upon plugging (\ref{eq:A Normalized Fixed Point Iteration-7}) into (\ref{eq:A Normalized Fixed Point Iteration-8}), we arrive at
\begin{align}\label{eq:A Normalized Fixed Point Iteration-10}
\frac{1}{\gamma^{*}}=\max_{k}\frac{\tilde{T}_{k}^{a_{k}^{*}}(\pmb{p}^{*})}{p_{k}^{*}}=\|\tilde{T}(\pmb{p}^{*})\|_{\infty}^{\bar{\pmb{p}}}.
\end{align}
Upon plugging (\ref{eq:A Normalized Fixed Point Iteration-10}) into (\ref{eq:A Normalized Fixed Point Iteration-7}), we can obtain (\ref{eq:A Normalized Fixed Point Iteration-4}).
\end{IEEEproof}
Based on the fixed point equation (\ref{eq:A Normalized Fixed Point Iteration-4}), we propose an NFP algorithm to solve problem (P)
(See Table \ref{table of normalized fixed point_simo}).
\begin{table}[htbp]
 \caption{NFP Algorithm for UL SIMO Cellular Networks}
 \label{table of normalized fixed point_simo}
 \large
\begin{tabular}{p{460pt}}
\hline
Initialization: pick random positive power vector $\pmb{p}(0)$.
\\
Loop $t$:
\\ 1) Compute the optimal beamforming vector $\{\hat{\pmb{u}}_{n,k}(t)\}$:
\\ 2) Compute BS association: $a_k(t) \leftarrow \tilde{A}_k(\pmb{p}(t)), \ \forall\; k $.
\\ 3) Update power: $ \pmb{p}(t+1) \leftarrow  \tilde{T}( \pmb{p}(t))$ ;
\\ 4) Normalize: $ \pmb{p}(t+1) \leftarrow \frac{\pmb{p}(t+1)}{ \| \pmb{p}(t+1) \|_{\infty}^{\bar{p}} }  $ ,  where $ \| \pmb{p}(t+1) \|_{\infty}^{\bar{p}} = \max_k \frac{ p_k(t+1) }{ \bar{p}_k} $.
\\ Iterate until convergence.
\\
\hline
\end{tabular}
\end{table}
The convergence property of this algorithm is given in the following result.
\begin{theorem}\label{thm of convergence_simo}
Suppose $(\bm p^*,  \{\pmb{u}_{n,k}^{*}\}_{k=1,\dots,K,\ n=1,\dots, N}, \bm a^*)$ is an optimal solution to problem (P).
Then the sequence $\{\bm p(t)\} $ generated by the NFP algorithm in Table \ref{table of normalized fixed point_simo} converges geometrically to $\bm p^*$, i.e.,
\begin{equation}\label{converg speed}
 \|\bm p(t) - \bm p^* \|_{\infty}^{\bar{p}} \leq C  \kappa^t,
\end{equation}
where $C>0,\ 0< \kappa < 1$ are constants that depend only on the problem data. 
\end{theorem}
Before proving Theorem 4, we introduce the following lemma.
\begin{lemma}\label{lemma of fixed point_simo_convave_mapping}
 The mapping $\tilde{T}(\pmb{p})$ is concave.
\end{lemma}
\begin{IEEEproof}[Proof of Lemma \ref{lemma of fixed point_simo_convave_mapping}]
For fixed $\pmb{p}$, $\tilde{T}_{k}^{n}(\pmb{p})$ is the minimum of a family of functions
\begin{displaymath}
\left\{ \hat{T}_{k}^{n}(\pmb{p},\pmb{u}_{n,k})=\frac{\sigma_{n}^{2}+\pmb{u}_{n,k}^{H}\sum_{j=1,j\not=k}^{K}p_{j}\pmb{h}_{n,j}\pmb{h}_{n,j}^{H}\pmb{u}_{n,k}}{\pmb{u}_{n,k}^{H}\pmb{h}_{n,k}\pmb{h}_{n,k}^{H}\pmb{u}_{n,k}}\right\}_{\substack{\|\pmb{u}_{n,k}\|\\=1}},
\end{displaymath}
thus $\tilde{T}_{k}^{n}(\pmb{p})$ is a concave function of $\pmb{p}$.  Furthermore, the function $\tilde{T}_{k}(\pmb{p})$ defined in (\ref{eq:uplink_assocation_miso_minimum_power_model_7}) is the  minimum of $N$ concave functions $\tilde{T}_{k}^{n}(\pmb{p}),n=1,\dots,N$, hence $\tilde{T}(\pmb{p})=(\tilde{T}_{1}(\pmb{p}),\tilde{T}_{2}(\pmb{p}),\dots, \tilde{T}_{K}(\pmb{p}))$ is a concave function.  Consequently,  the mapping $\tilde{T}(\pmb{p})$ is a concave mapping.
\end{IEEEproof}
\begin{IEEEproof}[Proof of Theorem \ref{thm of convergence_simo}]
See Appendix \ref{sec:proof_of_theorem_4}.
\end{IEEEproof}

\begin{remark}\label{pseudo_polynomial_time_simo}
 Theorem 4 implies the pseudo-polynomial time solvability of problem \eqref{max min}.
Without loss of generality, we can assume $\sigma_n^2 = 1$ \cite{schubert2004}, which does not change problem \eqref{max min} and the NFP algorithm in Table \ref{table of normalized fixed point_simo}.
Based on Cauchy-Schwarz inequality, we have $\|\pmb{u}_{n,k}^{H}\pmb{h}_{n,j}\|^{2}\leq \|\pmb{h}_{n,j}\|^{2}$.  Hence, it is easy to verify that $ \kappa \leq 1 - 1/( K G \cdot \mathrm{SNR} +1 ) $, where $\mathrm{SNR} = \max_{k} \bar{p}_k$ and $G = \max_{n,k} \{ \|\pmb{h}_{nk}\|^{2} \} $. 
 To achieve an $\epsilon$-optimal solution, the NFP algorithm in Table \ref{table of normalized fixed point_simo} takes $T \leq \frac{\log(1/\epsilon)}{ \log( 1/\kappa ) } \leq \log(1/\epsilon) (K G \cdot \mathrm{SNR}+1)$ iterations, where we have used the property  $-\log(1-x)>x, \textrm{when}~ x<1$.
 Since $K G \cdot \mathrm{SNR}$ is polynomial in the input parameters $K, \{\bar{p}_k\}$ and $\{ \|\pmb{h}_{nk}\|^{2}\} $, we obtain the pseudo-polynomial time solvability of problem \eqref{max min}.    Note that to prove the polynomial time solvability, we need to show that $T$ is upper bounded by a polynomial function of $K$, $\{\log p_{k}\}$ and $\{\log\|\pmb{h}_{nk}\|^{2}\}$.
\end{remark}
\begin{remark}\label{pseudo_polynomial_time_simo_fixed_bs_association}
With fixed BS association, problem (P) becomes a joint beamforming and power allocation problem in an SIMO I-MAC. We can adapt the NFP algorithm in Table  \ref{table of normalized fixed point_simo} to solve this simplified problem (assuming fixed BS association $\bm a$): replacing $\hat{\pmb{u}}_{n,k}(t)$ with $\hat{\pmb{u}}_{a_k,k}(t)$, skipping step 2, and replacing $\tilde{T}( \pmb{p}(t))$ with $\tilde{T}^{\pmb{a}}(\pmb{p})\triangleq(\tilde{T}_{1}^{a_{1}}(\pmb{p}),\tilde{T}_{2}^{a_{2}}(\pmb{p}),\dots, \tilde{T}_{K}^{a_{K}}(\pmb{p}))$. Using a similar argument, we can prove that this simplified algorithm also converges to the global optima geometrically.
\end{remark}
\section{Simulation Results}\label{sec:Simulation_Results_miso_max_min}
In this section, numerical results are provided to demonstrate the performance of the proposed algorithms.   We consider both  homogeneous networks (HomoNets) and  heterogeneous networks (HetNets).
For HomoNets, each macro cell contains one  macro BS in the center and the distance between adjacent macro BSs is 1000m.
For HetNets,  we assume that each macro cell contains one  macro BS in the center and there are 3 pico BSs randomly placed in each macro cell.
There are $K$ users with the same power budget $\bar{p}_k = P_{\max}$ in the network and we consider two user distributions: in ``Uniform'', users are uniformly distributed in the network area;
in ``Congested'', $K/4 $ users are placed randomly in one macro cell, while other users are uniformly distributed in the network area.
For SISO cellular networks, $g_{nk} =S_{nk}(200/d_{nk})^{3.7} $ where $d_{nk}$ is the distance from user $k$
to BS $n$ and  $10 \log_{10} S_{n,k} \sim \mathcal{N}(0,64)$ models the shadowing effect.    For SIMO cellular networks, the number of antennas at each BS is set to be the same as  $M_{1}=\cdots=M_{N}=4$ and the channel coefficients
between user $k$ and BS $n$ are modeled  as zero mean circularly symmetric complex Gaussian vector with $S_{nk}(200/d_{nk})^{3.7}$ being the variance for both real and imaginary dimensions.   Suppose the noise power is $\sigma = 1$, and define the signal to noise ratio
as $\text{SNR} = 10 \log_{10}(P_{\max})$.

\subsection{Comparison of Average Computation Time }\label{sec:Averaged_Computation_Time_Comparison}
Firstly, the average computation time is considered as  the efficiency indicator of the three different algorithms.
We perform the numerical experiments in a PC  with a Pentium G2030 3GHz CPU, 4GB RAM and Matlab R2014a.
\begin{table}[htbp]
\caption{Comparison of Average Computation Time Used by Different Algorithms for a SISO Scenario}
\label{table of computation_time_for_SISO}
\centering
\begin{tabular}{|c|c|c|c|c|c|c|c|}
\hline
SNR (dB)& 0 & 5 & 10 & 15 & 20 & 25 & 30\\
\hline
Time (s) BS-LP &70.8511 &  87.7384  & 116.8736  &  122.2134 &  102.6065 &  81.3555 &  70.0839\\
\hline
Time (s) BS-FP & 0.0112  &   0.0121  &  0.0139   & 0.0138 &    0.0204 &   0.0334 &    0.0519\\
\hline
Time (s) NFP & 0.0004    &  0.0005  &    0.0007  &     0.0009   &   0.0015    &  0.0021   &   0.0026\\
\hline
\end{tabular}
\end{table}

\begin{table}[htbp]
\caption{Comparison of Average Computation Time Used by Different Algorithms for a SIMO Scenario}
\label{table of computation_time_for_SIMO}
\centering
\begin{tabular}{|c|c|c|c|c|c|c|c|}
\hline
SNR (dB)& 0 & 5 & 10 & 15 & 20 & 25 & 30\\
\hline
Time (s) BS-SDP &61.0384 &  54.5044  & 68.0620 &  66.6358 &  70.2120  &69.8054  &  89.1961\\
\hline
Time (s) BS-FP &  0.2388  &  0.2022  &  0.4163  &  0.6571  & 1.1019  &   2.0289  &  4.2315\\
\hline
Time (s) NFP & 0.0207   & 0.0188   & 0.0233   & 0.0228 &  0.0231&  0.0232  &  0.0247\\
\hline
\end{tabular}
\end{table}

For the SISO scenario, we consider a HetNet  that consists of 10 hexagon macro cells.   There are 3 pico BSs randomly placed in each macro cell,  thus in total there are $N = 40$ BSs.  There are $K = 80$ users  uniformly distributed in the network area.      For BS-LP algorithm, the LP subproblem is solved by ``linprog'' function in Matlab with simplex method.    The average computation time is obtained by averaging over $500$ monte carlo runs and is listed in Table \ref{table of computation_time_for_SISO} and the stopping criterion is $\|\pmb{p}(t+1)-\pmb{p}(t)\| \leq \epsilon$, where $\epsilon=10^{-6}P_{\max}\sqrt{K}$.  As we can see from Table \ref{table of computation_time_for_SISO},  the NFP algorithm is at least 26000 times faster than BS-LP algorithm  and BS-FP algorithm is at least 1300 times faster than BS-LP algorithm for all considered SNR values.

For the SIMO scenario, we consider a HomoNet  that consists of 3 hexagon macro cells.  There are $K = 10$ users  uniformly distributed in the network area.
 For BS-SDP algorithm, the SDP subproblem is solved by CVX 2.1.   The average computation time is obtained by averaging over $500$ monte carlo runs and is listed in Table \ref{table of computation_time_for_SIMO} and the stopping criterion is $\|\pmb{p}(t+1)-\pmb{p}(t)\| \leq \epsilon$, where $\epsilon=10^{-6}P_{\max}\sqrt{K}$.  As we can see from Table \ref{table of computation_time_for_SIMO},
the NFP algorithm is at least 2800 times faster than BS-SDP algorithm for any SNR and BS-FP algorithm can be 21 to 260 times faster than BS-SDP algorithm depending on SNR.  Due to the high efficiency of both the BS-FP and the NFP algorithms, we  only investigate the performance of these two algorithms below.

\subsection{Comparison of Number of Iterations}\label{sec:Comparison of Number of Iteration}
The simulation scenarios  in the last subsection are limited to small size networks, as the running time required for {\color{black}BS-LP and BS-SDP algorithms } increases substantially with increasing number of  BSs and users.
In this subsection, we consider the scenarios with many more users and BSs than the scenarios considered in the last subsection to  further evaluate the performance of
the BS-FP and the NFP algorithms.  In particular, we consider a HetNet  that consists of 25 hexagon macro cells, each containing one macro BS in the center.  There are 3 pico BSs randomly placed in each macro cell, thus in total there are $N = 100$ BSs.   Furthermore, there are $K = 160$ users.
When only the FP algorithm is considered, it has similar computation complexity with one iteration of the NFP algorithm.  Hence, the biggest difference of the BS-FP and the NFP algorithms comes from the binary search invoked in BS-FP.
We will show that the binary search makes the BS-FP algorithm much slower than the NFP algorithm in terms of number of iterations.

Fig.\ 1 depicts the CDF (Cumulative Distribution Function) of the number of iterations needed in the context of SISO cellular networks for the following three algorithms to converge:
the BS-FP, the NFP and
the algorithm ``Oracle''  when $\textrm{SNR}=15$dB. 
In the algorithm ``Oracle'', we fix the BS association to be the optimal one $\bm a$, and compute the optimal power allocation
by the following procedure (proposed in \cite{TanChiang2009})
\begin{equation}\label{fix power update}
p_k(t+1) \leftarrow \frac{ T_k^{a_{k}}(\bm p(t))}{ \| T^{\pmb{a}}(\bm p(t))\|_{\infty}^{\bar{p}} },
\end{equation}
where  $T^{\pmb{a}}(\bm p)\triangleq [T_1^{a_{1}}(\bm p),T_2^{a_{2}}(\bm p),\dots, T_K^{a_{K}}(\bm p)]$.
A little surprisingly, the NFP algrorithm and the algorithm ``Oracle'' converge equally fast: they usually converge in 10$\sim$30 iterations. 
Due to the binary search step, BS-FP algorithms takes more than $150$ iterations in total to converge. 

Fig.\ 2 depicts the CDF  of the number of iterations needed in the context of SIMO cellular networks for the following three algorithms to converge:
the BS-FP, the NFP and
the algorithm ``Oracle'' when $\textrm{SNR}=10$dB. 
In the algorithm ``Oracle'', we fix the BS association to be the optimal one $\bm a$, and compute the optimal power allocation by the algorithm in Remark 3, i.e.
\begin{equation}\label{fix power update_SIMO}
p_k(t+1) \leftarrow \frac{ \tilde{T}_k^{a_{k}}(\bm p(t))}{ \| \tilde{T}^{\pmb{a}}(\bm p(t))\|_{\infty}^{\bar{p}} }.
\end{equation}
As mentioned in Remark 3, the above procedure also converges geometrically.
In  Fig.\ 2, it can be observed that the NFP algorithm and the algorithm ``Oracle'' converge equally fast: they usually converge in 20$\sim$40 iterations. 
Due to the binary search step, BS-FP algorithms takes more than $150$ iterations in total to converge.

\subsection{Comparison of Minimum SINR Achieved}\label{sec:Comparison of Minimum Rate Achieved}
{\color{black}In this subsection, the system performance is evaluated in terms of  achievable minimum SINR.}  The system configuration is the same as that in Subsection \ref{sec:Comparison of Number of Iteration}.

Fig.\ 3 compares the minimum SINR achieved by the BS-FP, the NFP and the ``max-SNR'' algorithm for  SISO cellular networks.
The ``max-SNR'' algorithm computes the BS association based on the maximum receive SNR, i.e. $a_k = \arg \max_n \{g_{nk}\bar{p}_{k} \} $.
For a fair comparison, the optimal power allocation corresponding to ``max-SNR'' algorithm is then computed by (\ref{fix power update}).
Each point in the figure is obtained by averaging over $500$ monte carlo runs.
The BS-FP and the NFP algorithms have similar performance in terms of the minimum rate.
For the setting ``Uniform'', the NFP algorithm outperforms ``max-SNR'' by approximately $70\%$ (when SNR$=35$dB);
 for ``Congested'', the NFP algorithm outperforms ``max-SNR'' by $400 \%$ (when SNR$=35$dB).

Fig.\ 4 compares the minimum SINR achieved by the BS-FP, the NFP and the ``max-SNR'' algorithms for SIMO cellular networks.
The ``max-SNR'' algorithm computes the BS association based on the maximum receive SNR, i.e. $a_k = \arg \max_n \{\|\pmb{h}_{nk}\|^{2}\bar{p}_{k} \} $.
For a fair comparison, the optimal power allocation corresponding to ``max-SNR'' algorithm is then computed by (\ref{fix power update_SIMO}).
Each point in the figure is obtained by averaging over $500$ monte carlo runs.
The BS-FP and the NFP algorithms almost have the  same performance in terms of the minimum rate.
For the setting ``Uniform'', the NFP algorithm outperforms ``max-SNR'' by approximately $35\%$ (when SNR$=25$dB);
 for ``Congested'', the NFP algorithm outperforms ``max-SNR'' by $200 \%$ (when SNR$=25$dB).

     \begin{figure}[ht]
{\includegraphics[width=1\linewidth]{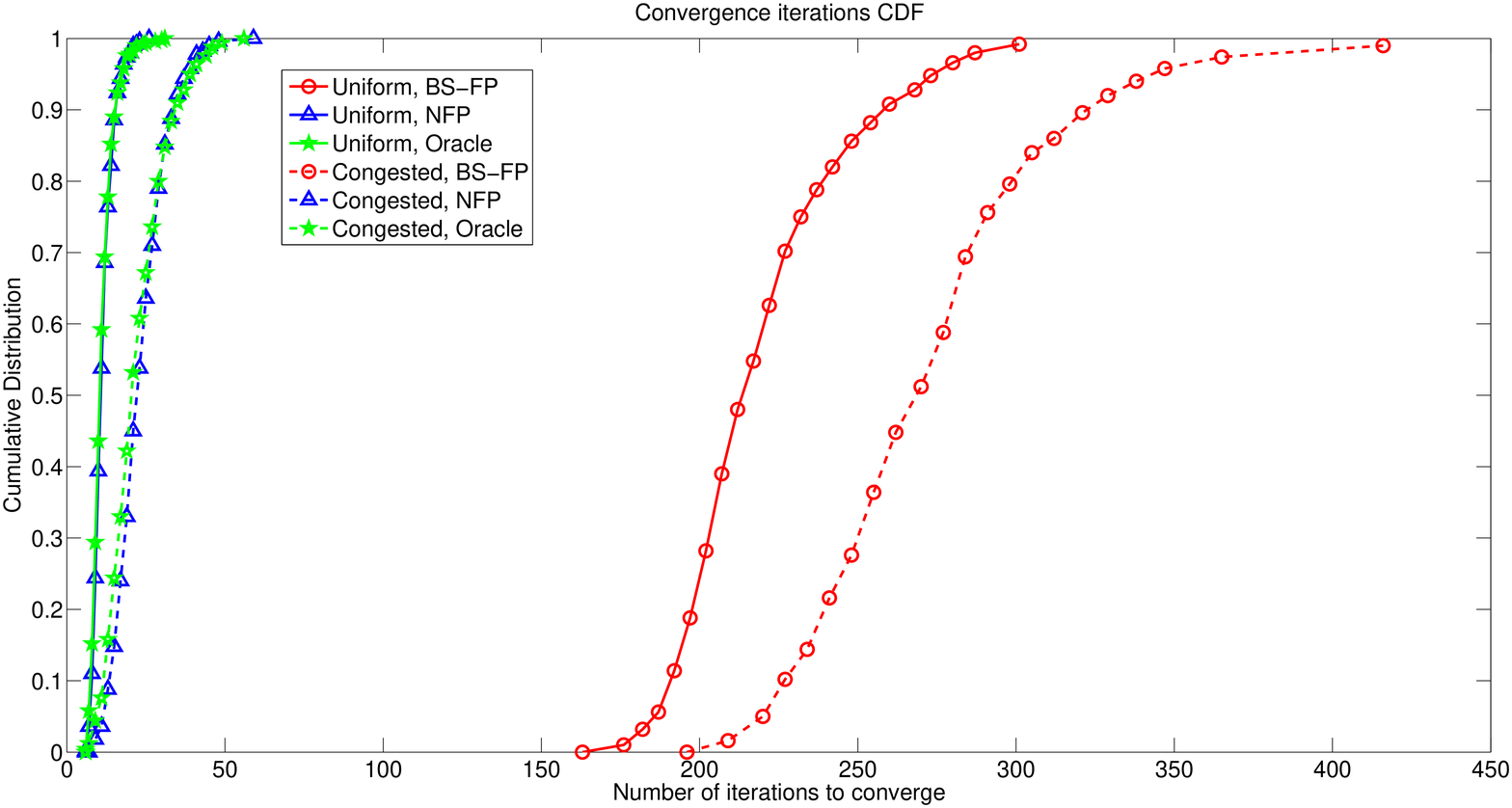} }
  \vspace{-.7cm}
   \caption{
    Distribution of the number of iterations required to converge for SISO cellular networks. $N=100$ BSs, $K=160$ users, $\text{SNR}=15$dB.
}  \label{figCDF} \vspace{-.3cm}
\end{figure}

     \begin{figure}[ht]
{\includegraphics[width=1\linewidth]{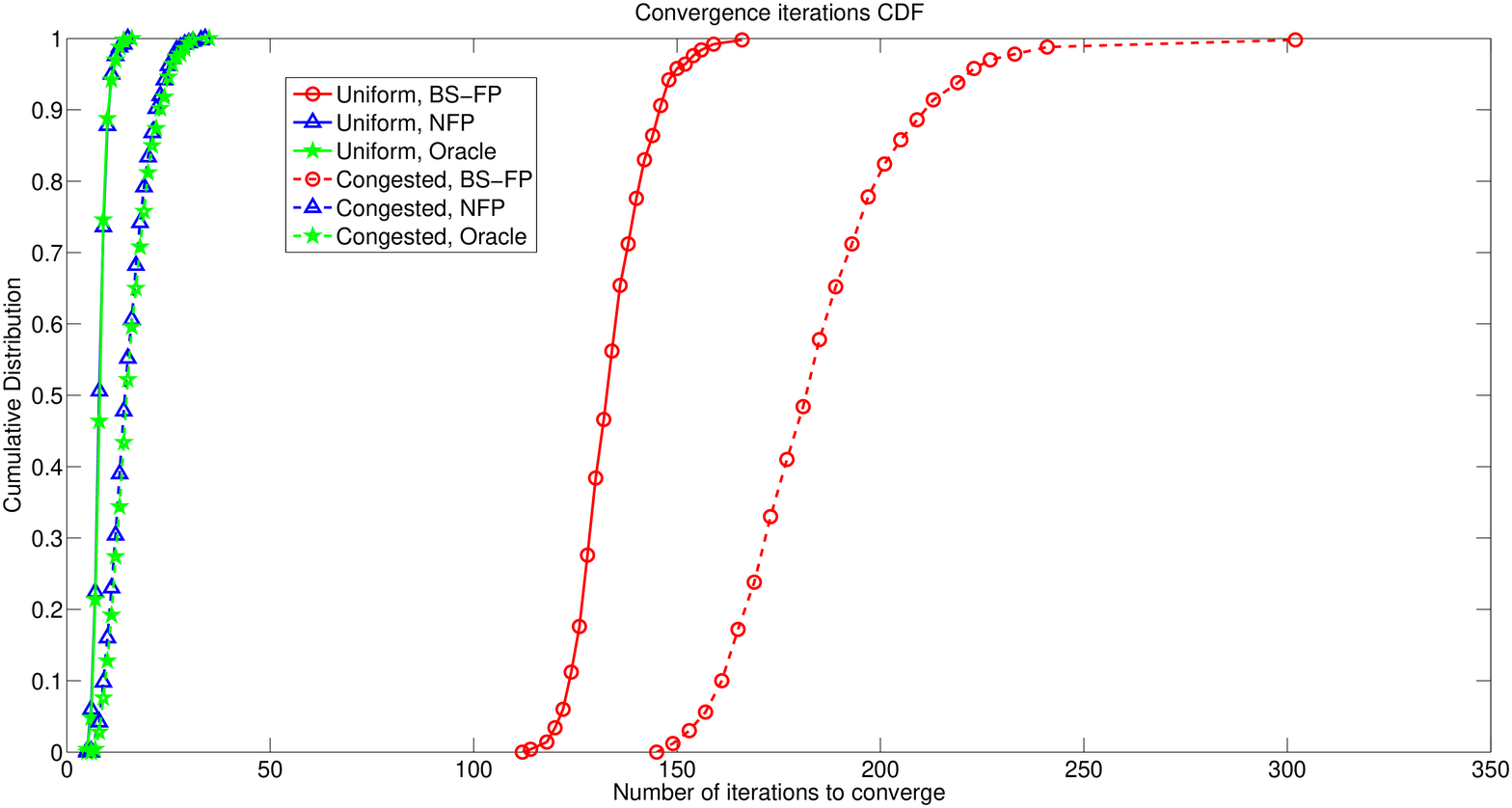} }
  \vspace{-.7cm}
   \caption{
    Distribution of the number of iterations required to converge for SIMO cellular networks. $N=100$ BSs, $K=160$ users, $\text{SNR}=10$dB.
}  \label{figCDF} \vspace{-.3cm}
\end{figure}

    \begin{figure}[ht]
{\includegraphics[width=1\linewidth]{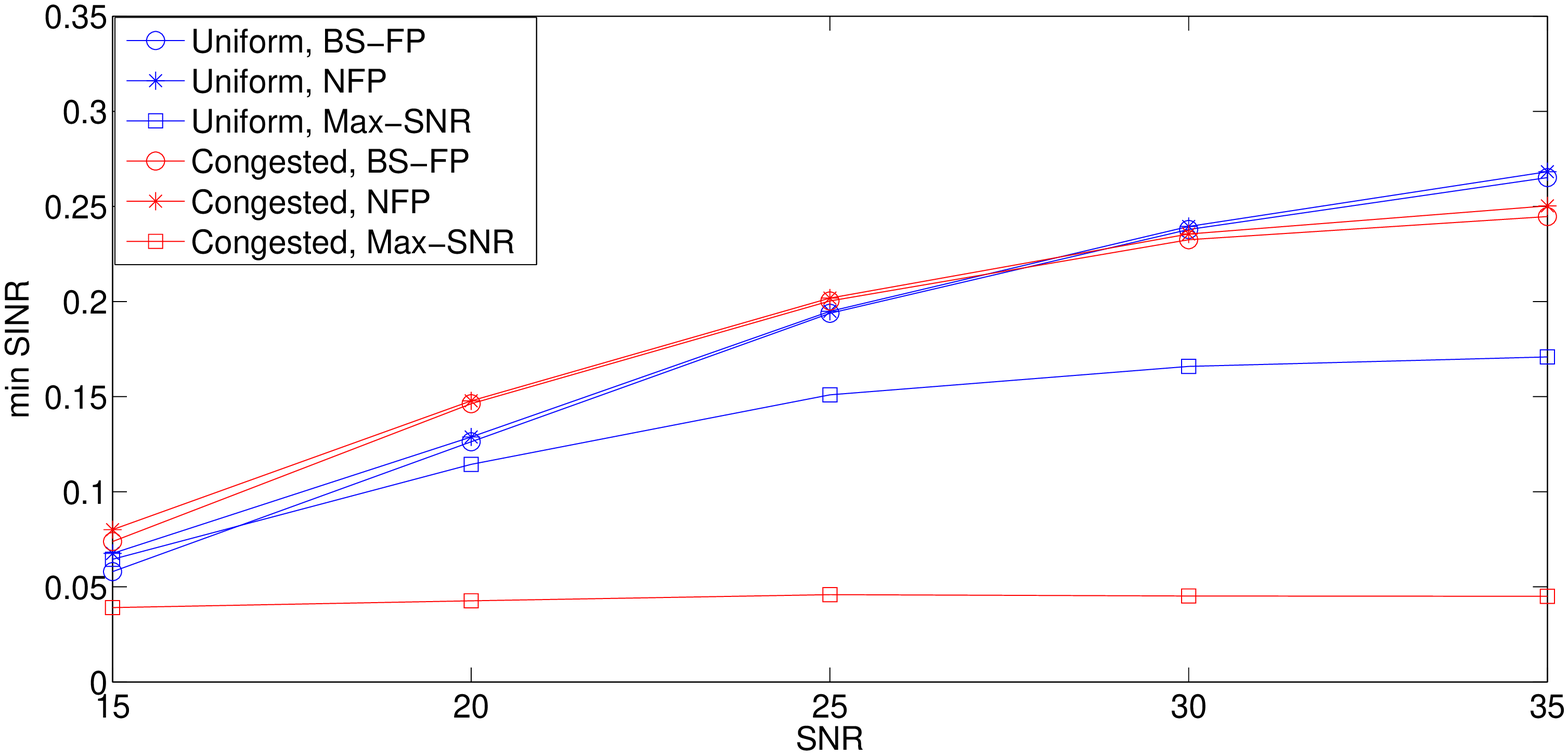} }
\vspace{-.7cm}
    \caption{
    Comparison of the minimum SINR achieved for SISO cellular networks. $N=100$ BSs, $K=160$ users.
}\label{figRate} \vspace{-.3cm}
\end{figure}

    \begin{figure}[ht]
{\includegraphics[width=1\linewidth]{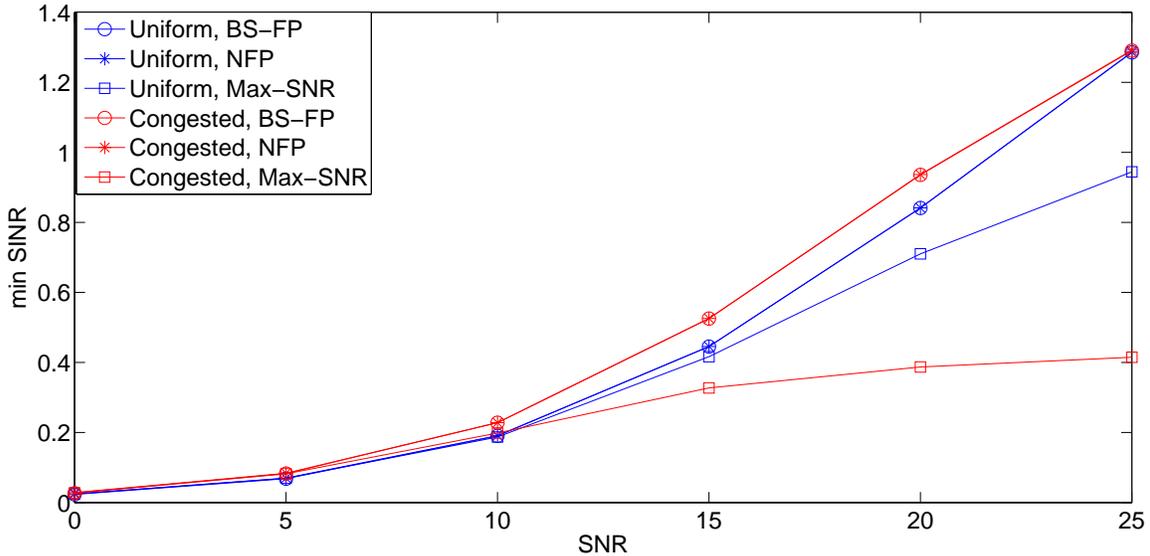} }
\vspace{-.7cm}
    \caption{
    Comparison of the minimum SINR achieved  for SIMO cellular networks. $N=100$ BSs, $K=160$ users.
}\label{figRate} \vspace{-.3cm}
\end{figure}

\section{Conclusions}\label{sec:Conclusions_miso_max_min}
In this paper, we investigate the joint BS association and beamforming problem for max-min fairness criterion in the context of UL SIMO cellular networks.
We prove the polynomial time solvability of the problem for both SISO and SIMO scenarios by transferring the original problem into a binary search method in conjunction with a series of QoS subproblems which can solved by LP for SISO or SDP for SIMO scenarios, yielding the so-called BS-LP and BS-SDP algorithms.
Furthermore, in order to avoid the computational complexity imposed by LP or SDP, we present a BS-FP algorithm where QoS subproblems are solved by a fixed point method.
Moreover, for the further reduction of computational complexity, we proposed a novel NFP algorithm which can directly solve the original problem without resorting
to the binary search.
We show that the NFP algorithm converges to the global optima at a geometric rate.
Though we are not able to prove that the NFP algorithm is a polynomial time algorithm, empirically it converges much faster than BS-FP and the provably polynomial time algorithm (BS-LP and BS-SDP). It is a theoretically interesting open question whether
the NFP algorithm is a polynomial time algorithm.
%
\begin{appendices}

\section{Proof of Proposition 1}\label{sec:proof_of_proposition_1}
\begin{IEEEproof}
We first prove the following fact:
if problem (P\textsubscript{SISO-QoS}) is feasible,  then its optimal power vector $\pmb{p}^{*}$ satisfies the fixed point equation
(\ref{min power pro, fixed point iteration}).   It can be seen that $\frac{p_{k}^{*}}{T_{k}(\pmb{p}^{*})}=\gamma$; otherwise, we can reduce the power $p_{k}^{*}$ to improve the objective function without violating all constraints. As $\pmb{p}^{*}$ satisfies the constraints of (P\textsubscript{SISO-QoS}),
 we have $\bar{\pmb{p}}_{k} \geq  p_{k}^{*}=\gamma T_{k}(\pmb{p}^{*}) $.  Consequently, $p_{k}^{*}=\min\{\gamma T_{k}(\pmb{p}^{*}), \bar{p}_{k}\}, \ \forall \ k$, which means that $\pmb{p}^{*}$ satisfies the fixed point equation of (\ref{min power pro, fixed point iteration}).

If $\gamma_{\textrm{ach}}=\gamma$, according to (\ref{min power pro, fixed point iteration_new_1}), we have $\textrm{SINR}_{k}=\frac{q_{k}}{ T_{k}(\pmb{q})} \geq \gamma, \forall k$.  Based on  (\ref{min power pro, fixed point iteration}), we have  $0\leq q_{k} \leq \bar{p}_{k}, \forall k$.  Hence $\pmb{q}$ satisfies the constraints of problem (P\textsubscript{SISO-QoS}), i.e. $\pmb{q}$ is a  feasible solution to (P\textsubscript{SISO-QoS}).   Assume $\pmb{p}^{*}$ is an optimal power vector to  (P\textsubscript{SISO-QoS}), then by the argument in the last paragraph $\pmb{p}^{*}$ satisfies (\ref{min power pro, fixed point iteration}).   As both $\pmb{q}$ and $\pmb{p}^{*}$ are fixed points of (\ref{min power pro, fixed point iteration}) and as mentioned earlier that according to \cite[Section V.B,Corollary 1]{Yates1997} $\pmb{q}$ is the unique fixed point of  (\ref{min power pro, fixed point iteration}),  we have $\pmb{q}= \pmb{p}^{*}$ and  $ (\bm q, \bm b) $ is an optimal solution to (P\textsubscript{SISO-QoS}).

If  $\gamma_{\textrm{ach}}<\gamma$, according to (\ref{min power pro, fixed point iteration_new_1}) there exists at least one $q_{k}$ satisfying $q_{k} < \gamma T_{k}(\pmb{q})$ and for this $q_{k}$, based on (\ref{min power pro, fixed point iteration}), we have $\bar{p}_{k}=q_{k} < \gamma T_{k}(\pmb{q})$.    Assume (P\textsubscript{SISO-QoS}) is feasible and its optimal power vector is
  $\pmb{p}^{*}$,  we have $ \pmb{p}^{*}=\gamma T_{k}(p^{*})\leq \bar{p}_{k}, \ \forall \ k$, thus $\pmb{q} \not= \pmb{p}^{*}$.  Therefore,  $\pmb{q}$ and $\pmb{p}^{*}$ are two distinct  fixed points of (\ref{min power pro, fixed point iteration}),  which contradicts the the fact that  (\ref{min power pro, fixed point iteration}) has a unique fixed point.  Hence,  (P\textsubscript{SISO-QoS}) is infeasible.
\end{IEEEproof}

\section{Proof of Lemma  3}\label{sec:proof_of_lemma_3-1}
\begin{IEEEproof}
We can prove this Lemma by following the argument in \cite{Ya-Feng-Liu-1}.

Denote
\begin{align}\label{eq:subeq7_qos}
 f_{n,k}(\pmb{p})= p_{k}\pmb{h}_{n,k}^{H}\pmb{M}_{n}^{-1}(\pmb{p})\pmb{h}_{n,k},\ k=1,\dots,K, \ n=1,\dots,N,
\end{align}
which is  a strictly increasing function on $p_{k}\geq 0$ and a decreasing function on $p_{j},j\not= k$ \cite[Lemma~3.1]{Ya-Feng-Liu-1}.

Suppose there are two distinct solutions  $\tilde{\pmb{p}}$ and  $\hat{\pmb{p}}$ satisfying Eq.(\ref{eq:subeq4-1_qos}), i.e.
  \begin{equation}\label{eq:subeq4-1-2_qos}
  \max_{n}f_{n,k}(\tilde{\pmb{p}})=   \max_{n}f_{n,k}(\hat{\pmb{p}})=\frac{\gamma}{1+\gamma }, \; k=1,\dots,K.
\end{equation}

Define a nonempty set $\mathcal{K}=\{k\in K \mid \tilde{p}_{k}/\hat{p}_{k}>1\}$ and $k_{0}=\arg \max_{k\in \mathcal{K}}\{\tilde{p}_{k}/\hat{p}_{k}\}$.  Define the vector $\pmb{\alpha}=[\alpha_{1},\cdots,\alpha_{K}]$, where $\alpha_{k}$ is given by
\begin{align}\label{eq:subeq8_qos}
\alpha_{k}=\left\{ \begin{array}{ll}
          \frac{\tilde{p}_{k_{0}}}{\hat{p}_{k_{0}}}>1,& \mbox{if $k \in  \mathcal{K} $};\\
          1, & \mbox{otherwise}  ,\end{array} \right.
\end{align}
Consequently, we have
\begin{align}\label{eq:subeq8-1_qos}
 f_{n,k_{0}}(\tilde{\pmb{p}})&= f_{n,k_{0}}(\tilde{p}_{k_{0}},\tilde{\pmb{p}}_{-k_{0}})\overset{\textrm{(i)}} {\geq}f_{n,k_{0}}(\tilde{p}_{k_{0}},\pmb{\alpha}_{-k_
 {0}}\circ \hat{\pmb{p}}_{-k_{0}})\nonumber\\
 & \overset{\textrm{(ii)}} {=} \alpha_{k_{0}}\hat{p}_{k_{0}}\pmb{h}_{n,k_{0}}^{H}(\sigma_{n}^{2}\pmb{I}+\alpha_{k_{0}}\sum_{j\in\mathcal{K}}\pmb{h}_{n,j}\pmb{h}_{n,j}^{H}\hat{p}_{j}+\nonumber\\
 &\sum_{j\not\in\mathcal{K}}\pmb{h}_{n,j}\pmb{h}_{n,j}^{H}\hat{p}_{j})^{-1}\pmb{h}_{n,k_{0}}\nonumber\\
 &\overset{\textrm{(iii)}} {>} \hat{p}_{k_{0}}\pmb{h}_{n,k_ {0}}^{H}(\sigma_{n}^{2}\pmb{I}+\sum_{j\in\mathcal{K}}\pmb{h}_{n,j}\pmb{h}_{n,j}^{H}\hat{p}_{j}++\nonumber\\
 &\sum_{j\not\in\mathcal{K}}\pmb{h}_{n,j}\pmb{h}_{n,j}^{H}\hat{p}_{j})^{-1}\pmb{h}_{n,k_ {0}}\nonumber\\
&=f_{n,k_{0}}(\hat{\pmb{p}}),
\end{align}
where the notation $\circ$ denotes the Hadamard product,  $\tilde{\pmb{p}}_{-k_{0}}$ is the power vector with $k_0$th element deleted and   $\hat{\pmb{p}}_{-k_{0}}$ as well as $\pmb{\alpha}_{-k_ {0}}$ are defined analogously.  Moreover, (i) is due to $\tilde{\pmb{p}}_{-k_{0}} \leq \pmb{\alpha}_{-k_
 {0}}\circ \hat{\pmb{p}}_{-k_{0}}$,  (ii) is due to Eq.(\ref{eq:subeq8_qos}), while (iii) is due to $\alpha_{k_{0}}\hat{p}_{j}>\hat{p}_{j}, j\in\mathcal{K}$.

Consequently, we have
  \begin{equation}\label{eq:subeq4-1-3_qos}
  \max_{n}f_{n,k_{0}}(\tilde{\pmb{p}})>   \max_{n}f_{n,k_{0}}(\hat{\pmb{p}})=\frac{\gamma}{1+\gamma },
\end{equation}
which contradicts the eq.(\ref{eq:subeq4-1-2_qos}),
hence eq.(\ref{eq:subeq4-1_qos}) has a unique fixed point.
\end{IEEEproof}

\section{Proof of Lemma  4}\label{sec:proof_of_lemma_3}
\begin{IEEEproof}
For a given $\gamma$, if (P\textsubscript{SIMO-QoS-1}) is infeasible, suppose $\tilde{\pmb{p}}$ is one optimal solution to (P\textsubscript{SIMO-QoS-2}),
then $\max_{n \in \{1,\dots, N\}}  \tilde{p}_{k} \pmb{h}_{n,k}^{H}\pmb{M}_{n}^{-1}(\tilde{\pmb{p}})\pmb{h}_{n,k}   \leq \frac{\gamma}{1+\gamma}, \forall k $.
 We must have $ \max_{n \in \{1,\dots, N\}}  \tilde{p}_{k} \pmb{h}_{n,k}^{H}\pmb{M}_{n}^{-1}(\tilde{\pmb{p}})\pmb{h}_{n,k}   \leq \frac{\gamma}{1+\gamma} $
for some $k$; otherwise $ \max_{n \in \{1,\dots, N\}}  \tilde{p}_{k} \pmb{h}_{n,k}^{H}\pmb{M}_{n}^{-1}(\tilde{\pmb{p}})\pmb{h}_{n,k}  = \frac{\gamma}{1+\gamma} , \forall k $, implying that
$\tilde{\bm p}$ is a feasible solution to \textrm{(P\textsubscript{SIMO-QoS-1})}, a contradiction.

If (P\textsubscript{SIMO-QoS-1}) is feasible, denote its optimal solution as $\hat{\pmb{p}}$.  According to Proposition \ref{prop: feasibility check_simo},
 $\hat{\pmb{p}}$ is the unique solution to (\ref{eq:subeq6-1_qos}).  Assume  $\tilde{\pmb{p}}$ is one solution to (P\textsubscript{SIMO-QoS-2}) with $\tilde{\pmb{p}} \not= \hat{\pmb{p}}$.  In this case, we have
    \begin{align}\label{eq:subeq7-1_qos}
\max_{n}f_{n,k}(\tilde{\pmb{p}} ) \leq  \frac{\gamma}{1+\gamma}, \ k=1,\dots,K.
\end{align}
and $\sum_{k=1}^{K} \tilde{p}_{k} \geq \sum_{k=1}^{K} \hat{p}_{k}$.

Define a nonempty set $\mathcal{K}=\{k\in K \mid \tilde{p}_{k}/\hat{p}_{k}>1\}$ and $k_{0}=\arg \max_{k\in \mathcal{K}}\{\tilde{p}_{k}/\hat{p}_{k}\}$.  Define the vector $\pmb{\alpha}=[\alpha_{1},\cdots,\alpha_{K}]$, where $\alpha_{k}$ is given by
\begin{align}\label{eq:subeq8-2_qos}
\alpha_{k}=\left\{ \begin{array}{ll}
          \frac{\tilde{p}_{k_{0}}}{\hat{p}_{k_{0}}}>1,& \mbox{if $k \in  \mathcal{K} $};\\
          1, & \mbox{otherwise}  ,\end{array} \right.
\end{align}
Consequently, we have
\begin{align}\label{eq:subeq8-3_qos}
 f_{n,k_{0}}(\tilde{\pmb{p}})&= f_{n,k_{0}}(\tilde{p}_{k_{0}},\tilde{\pmb{p}}_{-k_{0}})\geq f_{n,k_{0}}(\tilde{p}_{k_{0}},\pmb{\alpha}_{-k_
 {0}}\circ \hat{\pmb{p}}_{-k_{0}})\nonumber\\
 & = \alpha_{k_{0}}\hat{p}_{k_{0}}\pmb{h}_{n,k_{0}}^{H}(\sigma_{n}^{2}\pmb{I}+\alpha_{k_{0}}\sum_{j\in\mathcal{K}}\pmb{h}_{n,j}\pmb{h}_{n,j}^{H}\hat{p}_{j}++\nonumber\\
 &\sum_{j\not\in\mathcal{K}}\pmb{h}_{n,j}\pmb{h}_{n,j}^{H}\hat{p}_{j})^{-1}\pmb{h}_{n,k_{0}}\nonumber\\
 &>\hat{p}_{k_{0}}\pmb{h}_{n,k_ {0}}^{H}(\sigma_{n}^{2}\pmb{I}+\sum_{j\in\mathcal{K}}\pmb{h}_{n,j}\pmb{h}_{n,j}^{H}\hat{p}_{j}++\nonumber\\
 &\sum_{j\not\in\mathcal{K}}\pmb{h}_{n,j}\pmb{h}_{n,j}^{H}\hat{p}_{j})^{-1}\pmb{h}_{n,k_ {0}}\nonumber\\
&=f_{n,k_{0}}(\hat{\pmb{p}})
\end{align}
Consequently, we have
  \begin{equation}\label{eq:subeq4-1-3_qos}
  \max_{n}f_{n,k_{0}}(\tilde{\pmb{p}})>   \max_{n}f_{n,k_{0}}(\hat{\pmb{p}})=\frac{\gamma}{1+\gamma },
\end{equation}
which contradicts  Eq.(\ref{eq:subeq7-1_qos}).

Consequently, if problems   (P\textsubscript{SIMO-QoS-1}) is feasible, the  problems   (P\textsubscript{SIMO-QoS-1}) and (P\textsubscript{SIMO-QoS-2}) have the same solution.
\end{IEEEproof}

\section{Proof of Lemma  5}\label{sec:Appendix_canonical form of_MIMO_OSTBC_2}
\begin{IEEEproof}
In order to show that $\tilde{T}_{k}^{n}(\pmb{p})$ is a standard interference function, we need to show three properties:
\begin{enumerate}
 \item   Positivity: For $\pmb{p}\geq \pmb{0}$, $\tilde{T}_{k}^{n}(\pmb{p}) > 0$;
 \item   Monotonicity:  If $\pmb{p}\geq \pmb{p}^{'}$, then $\tilde{T}_{k}^{n}(\pmb{p}) \geq \tilde{T}_{k}^{n}(\pmb{p}^{'})$;
 \item   Scalability:  For any $\alpha>1$, $\alpha \tilde{T}_{k}^{n}(\pmb{p}) > \tilde{T}_{k}^{n}(\alpha \pmb{p})$.
\end{enumerate}

 1) is obvious; 2) can be obtained from \cite[Lemma 2 (c)]{Farrokh-Rashid-Farrokhi}.  In order to show the scalability, we have
\begin{align}\label{eq:uplink_assocation_miso_minimum_power_model_6}
\tilde{T}_{k}^{n}(\alpha \pmb{p})&=\min_{\|\pmb{u}_{n,k}\|=1} \left\{    \frac{\sigma_{a_{k}}^{2}+\alpha\pmb{u}_{n,k}^{H}\sum_{j=1,j\not=k}^{K}p_{j}\pmb{h}_{n,j}\pmb{h}_{n,j}^{H}\pmb{u}_{n,k}}{\pmb{u}_{n,k}^{H}\pmb{h}_{n,k}\pmb{h}_{n,k}^{H}\pmb{u}_{n,k}}    \right\}\nonumber\\
  &< \alpha \min_{\|\pmb{u}_{n,k}\|=1} \left\{    \frac{\sigma_{n}^{2}+\pmb{u}_{n,k}^{H}\sum_{j=1,j\not=k}^{K}p_{j}\pmb{h}_{n,j}\pmb{h}_{n,j}^{H}\pmb{u}_{n,k}}{\pmb{u}_{n,k}^{H}\pmb{h}_{n,k}\pmb{h}_{n,k}^{H}\pmb{u}_{n,k}}    \right\}\nonumber\\
  &=\alpha \tilde{T}_{k}^{n}(\pmb{p}).
\end{align}
Hence  $\tilde{T}_{k}^{n}(\pmb{p})$ is a standard interference function.
\end{IEEEproof}

\section{Proof of Theorem \ref{thm of convergence_simo}}\label{sec:proof_of_theorem_4}
\begin{IEEEproof}
According to Lemma \ref{lemma of fixed point_simo}, $\pmb{p}^{*}$ is a fixed point of (\ref{eq:A Normalized Fixed Point Iteration-4}).  According to  the concave Perron-Frobenius theory\cite[Theorem 1]{krause2001concave}, we know that (\ref{eq:A Normalized Fixed Point Iteration-4})  has a unique fixed point and the NFP algorithm in Table \ref{table of normalized fixed point_simo} converges to this fixed point.  Hence, the NFP algorithm in Table \ref{table of normalized fixed point_simo} converges to $\pmb{p}^{*}$.

Define $U$ as the set of power vectors $\pmb{p}$ with $\|\pmb{p}\|_{\infty}^{\bar{\pmb{p}}}=1$.  It can be verified that
\begin{align}\label{eq:A Normalized Fixed Point Iteration-12}
A_{k}\leq \tilde{T}_{k}(\pmb{p}) \leq B_{k},\ \forall \ \pmb{p} \in U,
\end{align}
where $A_{k}=\min_{n}\min_{\{\|\pmb{u}_{n,k}\|=1\}} \frac{\sigma_{n}^{2}}{\pmb{u}_{n,k}^{H}\pmb{h}_{n,k}\pmb{h}_{n,k}^{H}\pmb{u}_{n,k}}=\min_{n} \frac{\sigma_{n}^{2}}{\|\pmb{h}_{n,k}\|^{2}}$ (note that  the second equality is based on the  Cauchy-Schwarz inequality of  $\|\pmb{u}_{n,k}^{H}\pmb{h}_{n,j}\|^{2}\leq \|\pmb{h}_{n,j}\|^{2}$),  and
$B_{k}=\tilde{T}_{k}(\bar{\pmb{p}})=\\ \min_{n} \min_{\{\|\pmb{u}_{n,k}\|=1\}}   \frac{\sigma_{n}^{2}+\pmb{u}_{n,k}^{H}\sum_{j=1,j\not=k}^{K}\pmb{h}_{n,j}\pmb{h}_{n,j}^{H}\pmb{u}_{n,k}\bar{p}_{j}}{\pmb{u}_{n,k}^{H}\pmb{h}_{n,k}\pmb{h}_{n,k}^{H}\pmb{u}_{n,k}}=\min_{n}\frac{1}{\bar{p}_{k}\pmb{h}_{n,k}\pmb{M}_{n}^{-1}(\bar{\pmb{p}})\pmb{h}_{n,k}}-1$,
both of which are constants that only depend on the problem data.    Based on (\ref{eq:A Normalized Fixed Point Iteration-12}), we have
 \begin{align}\label{eq:A Normalized Fixed Point Iteration-13}
(1-\kappa)\pmb{e} \leq \tilde{T}(\pmb{p}) \leq \pmb{e}, \ \forall  \ \pmb{p} \in U,
\end{align}
where $\kappa=1-\min_{k}\frac{A_{k}}{B_{k}} \in (0,1)$ and $\pmb{e}=(B_{1},B_{2},\dots, B_{K})>0$.
According to the concave Perron-Frobenius Theory \cite[Lemma 3, Theorem]{krause1994relative}, the NFP algorithm in Table \ref{table of normalized fixed point_simo} converges geometrically at the rate $\kappa$.
\end{IEEEproof}

\end{appendices}


\bibliographystyle{IEEEbib}
\bibliography{refsUplinkBSA}

\end{document}